\documentclass[%
preprint,
 amsmath,amssymb,
 aps,
]{revtex4-2}

\usepackage{graphicx}
\usepackage{dcolumn}
\usepackage{bm}
\usepackage [english]{babel}
\usepackage [autostyle, english = american]{csquotes}

\begin{document}

This manuscript has been authored by UT-Battelle, LLC under Contract No. DE-AC05-00OR22725 with the U.S. Department of Energy. The United States Government retains and the publisher, by accepting the article for publication, acknowledges that the United States Government retains a non-exclusive, paid-up, irrevocable, world-wide license to publish or reproduce the published form of this manuscript, or allow others to do so, for United States Government purposes. The Department of Energy will provide public access to these results of federally sponsored research in accordance with the DOE Public Access Plan(http://energy.gov/downloads/doe-public-access-plan).

\title{Microscopic view of heat capacity of matter: solid, liquid, and gas}

\author{Jaeyun Moon}
\email{To whom correspondence should be addressed; E-mail: moonj@ornl.gov}
 \affiliation{Materials Science and Technology Division,\\ Oak Ridge National Laboratory, Oak Ridge, Tennessee 37831, USA}

\author{Simon Th\'ebaud}
 \affiliation{Materials Science and Technology Division,\\ Oak Ridge National Laboratory, Oak Ridge, Tennessee 37831, USA}
\author{Lucas Lindsay}
 \affiliation{Materials Science and Technology Division,\\ Oak Ridge National Laboratory, Oak Ridge, Tennessee 37831, USA}
 
\author{Takeshi Egami}
\affiliation{Materials Science and Technology Division,\\ Oak Ridge National Laboratory, Oak Ridge, Tennessee 37831, USA\\
 Department of Materials Science and Engineering,\\ University of Tennessee, Knoxville, Tennessee 37996, USA\\
 Department of Physics and Astronomy,\\ University of Tennessee, Knoxville, Tennessee 37996, USA}

\date{\today}
\clearpage

\begin{abstract}
Understanding thermodynamics in liquids at the atomic level is challenging because of strong atomic interactions and lack of symmetry. Recent prior theoretical works have focused on describing heat capacity of liquids in terms of phonon-like excitations but often rely on fitting parameters and ad hoc assumptions. In this work, we perform microscopic analysis on instantaneous normal modes and velocity autocorrelations on molecular dynamics simulations of single element systems over wide ranges of temperature (up to 10\textsuperscript{8} K) and pressure (up to 1 TPa). Our results demonstrate that heat capacity of liquids can be described by a combination of both solid-like and gas-like degrees of freedom, leading to a unified framework to describe heat capacity of all three phases of matter: solid, liquid, and gas.  
\end{abstract}


\maketitle
\clearpage

In various applications of liquids from nuclear reactors \cite{williams_assessment_2006-1} to thermal energy storage devices \cite{li_sensible_2016}, performance and efficiency of such device are directly controlled by heat capacity. However, engineering and customization of heat capacity of liquids are challenging due to a general lack of microscopic understanding of the thermodynamics of liquids. Liquids have disordered structures lacking spatial periodicity, yet their physical densities are similar to solids. In gases, atoms are similarly disordered; however, liquids are strongly interacting and dynamically correlated. These characteristics make rigorous theoretical treatment of liquid thermodynamics a daunting challenge. As such, the physics of liquids has been historically studied starting from more established gas and solid perspectives. 

Considering liquid and gas states as a unified concept dates back two centuries to Cagniard de la Tour \cite{de_la_tour_expose_1822}, Faraday, Whewell, Dumas \cite{faraday_xvii_1823, faraday_selected_1971}, Mendeleev \cite{mendelejeff_ueber_1861}, Andrews \cite{andrews_bakerian_1869, tait_scientific_1889}, and van der Waals \cite{van_der_waals_over_1873}. The work by van der Waals also motivated development of hard-sphere paradigm that is widely used to study soft matter, granular materials, gases, and liquids \cite{chandler_van_1983, dyre_simple_2016}. On the other hand, many have also suggested that liquids can be modeled by atoms as organized in crystals  \cite{mie_zur_1903,mott_resistance_1934,lennard-jones_critical_1937,granato_specific_2002}. Frenkel considered liquids and solids as a continuity \cite{frenkel_uber_1926, frenkel_continuity_1935, frenkel_liquid_1937,frenkel_kinetic_1947} and unified these phases of matter as ``condensed bodies". Frenkel proposed that when the characteristic time of interest is less than $\tau = \eta/\mathcal{N}$ where $\eta$ is viscosity and $\mathcal{N}$ is rigidity modulus, materials behave like solid, and otherwise like liquid \cite{frenkel_liquid_1937}. The use of the term ``condensed matter" has been since popularized by  Anderson and Heine to include liquid physics in the 1960s. Frenkel's seminal ideas have led to many recent works considering thermodynamics of liquids from phonon quasi-particles as in solids \cite{trachenko_heat_2011,bolmatov_phonon_2012,andritsos_heat_2013,trachenko_collective_2016,wang_direct_2017,yang_emergence_2017,brazhkin_liquid-like_2018,tomiyoshi_heat_2019,khusnutdinoff_collective_2020,Kryuchkov_universal_2020,zaccone_universal_2021,baggioli_explaining_2021}.  

Motivated by these historical accounts studying liquids from either gas or solid perspectives, we consider here liquid thermodynamics and thermal properties as a subset of a more general, unified framework that includes all classical phases of matter simultaneously: solids, liquids, and gases. To obtain microscopic understanding of thermodynamics and thermal properties of matter, it is imperative to characterize the effective heat carriers and understand their nature. We build conclusions of effective heat carriers in liquids from systems over a very broad range of thermodynamic states rather than from a limited temperature and pressure ranges typically reported in the literature \cite{bolmatov_phonon_2012, baggioli_explaining_2021,zaccone_universal_2021,tomiyoshi_heat_2019,Kryuchkov_universal_2020}.

Here we focus our discussion on heat carriers related to atomic motion. In solids, they are phonons characterized by normal modes of the lattice vibration. Phonon energies ($\hbar \omega$) and their spectral distribution, $g(\omega)$, i.e., the density of states, are typically calculated by diagonalizing the dynamical matrices built from atomic positions and force constants or using velocity autocorrelation functions (VACF($\omega$)) from molecular dynamics - both methods yield equivalent results under the harmonic approximation. In this approximation, total energy and constant volume heat capacity can be expressed in terms of phonons by $E = \int d\omega \hbar \omega (n + \frac{1}{2}) g(\omega)$ and $C_V = (dE/dT)_V = \int d\omega k_B (\frac{\hbar \omega}{2k_BT})^2 \sinh^{-2}(\hbar \omega/2k_BT) g(\omega)$, where $n$ is the Bose-Einstein distribution, and corrections can be further made to include phonon anharmonicity \cite{togo_first_2015}. In the classical limit, these become $E = 3Nk_BT$ and $C_V = 3Nk_B$, also known as the Dulong-Petit law of specific heat. Quadratic potential and kinetic energies lead to equipartition of energy with $\frac{1}{2}k_BT$ per degree of freedom. For gases where potential interactions are weak compared to thermal energy, an atomic particle picture is used to describe thermodynamics and thermal properties. Diagonalizing dynamical matrices of a gas state is rarely used even for dense gases where potential interactions are relevant, but VACF($\omega$) yields well-known Lorentzian lineshapes characterizing independent uncorrelated atomic collisions. For monatomic dilute gases, in which classical approximations are typically acceptable, total energy is equivalent to the total kinetic energy, $E = E_{KE} = \sum_i \frac{1}{2}m|\boldsymbol{v}_i|^2 = \frac{3}{2}Nk_BT$, where the sum is over all atoms and the corresponding specific heat is $C_V = \frac{3}{2}Nk_B$. 

These solid and gas limits ($C_V = 3Nk_B$ vs. $\frac{3}{2}Nk_B$) suggest that the effective heat carriers in liquids may potentially be described as an intermediate state with a varying degree of importance of potential interactions. In contrast, widely used heat capacity theories of liquids typically rely on phonon quasi-particles alone \cite{trachenko_heat_2011,bolmatov_phonon_2012,andritsos_heat_2013,trachenko_collective_2016,wang_direct_2017,yang_emergence_2017,brazhkin_liquid-like_2018,tomiyoshi_heat_2019,khusnutdinoff_collective_2020,Kryuchkov_universal_2020,zaccone_universal_2021,baggioli_explaining_2021} and describe the decrease in heat capacity of liquids with increase in temperature solely due to a disappearance of particular phonon populations \cite{zaccone_universal_2021,baggioli_explaining_2021}. Thus, the full translational atomic degrees of freedom ($3N$) are not taken into account \cite{zaccone_universal_2021,baggioli_explaining_2021}. In addition, these theories are based on assumptions of Debye densities of states \cite{bolmatov_phonon_2012} or Gaussian densities of states of normal modes at high frequencies with free fitting parameters \cite{zaccone_universal_2021,baggioli_explaining_2021}.     

In this work, we systematically characterize effective heat carriers of monatomic systems (argon, silicon, and iron) under constant volume from solid to gas through both instantaneous normal mode analysis and spectral velocity autocorrelation functions over a wide range of temperatures from 1 to 10\textsuperscript{8} K. Based on these microscopic calculations, we define general instability parameters (IP\textsubscript{1} and IP\textsubscript{2}) to describe the `gasness' of a system, and we draw connection between seemingly different instantaneous normal mode distributions and spectral velocity autocorrelation functions in liquid and gas states. We further demonstrate agreement between constant volume specific heat predicted from IPs and independent calculations using molecular dynamics (MD). These results pave the way towards a unified approach to thermodynamics of matter and thermal properties from solid to gas and provide new insights into a microscopic view of liquid thermodynamics. 

Interatomic interactions are described by Lennard-Jones (argon) \cite{jones_determination_1924, jones_determination_1924-1, rahman_molecular_1976}, Stillinger-Weber (silicon) \cite{stillinger_computer_1985}, and modified Johnson potentials (iron) \cite{srolovitz_structural_1981, levashov_equipartition_2008}. We focus our scope on specific heat over a wide range of temperatures maintaining a constant volume determined by the equilibrium density, $\rho_0$, at 1 K and 1 bar for each system to minimize the effect of anharmonicity. Effects of volumetric expansion and the role of anharmonicity may be addressed in future works. Detailed simulation procedures are discussed in Supplementary Materials. 

We perform molecular dynamics (MD) simulations using Large-scale Atomic/ Molecular Massively Parallel Simulator (LAMMPS) \cite{plimpton_fast_1995} to (1) generate equilibrated atomic structures at a given temperature, (2) compute spectral velocity autocorrelation functions, and (3) calculate specific heats. We further use the equilibrated atomic structures generated from MD to perform lattice dynamics calculations (GULP \cite{gale_gulp:_1997} and in-house codes) to obtain eigenfrequencies of the $\Gamma$ point dynamical matrices with the entire domain considered as a unit cell. For better statistics, 10 structures at each temperature in MD simulations were used for the lattice dynamics calculations. Three independent MD simulations using different initial velocities at each temperature were done for specific heat calculations.

The same procedures were additionally applied to structures at different densities to characterize the sensitivity to density and anharmonicity: $0.9 \rho_0$ and $0.8 \rho_0$ for argon, $1.1 \rho_0$ for silicon, and $0.9 \rho_0$ for iron. For silicon, due to a large relative increase in coordination numbers from 4 to more than 6 upon melting, reducing density to $0.9 \rho_0$ and $0.8 \rho_0$ led to segregation of atoms and large empty space in the simulation domains for some liquid temperatures. Thus, we instead chose a more dense $\rho = 1.1 \rho_0$ for silicon. 

\begin{figure}[h!]
	\centering
	\includegraphics[width=0.49\linewidth]{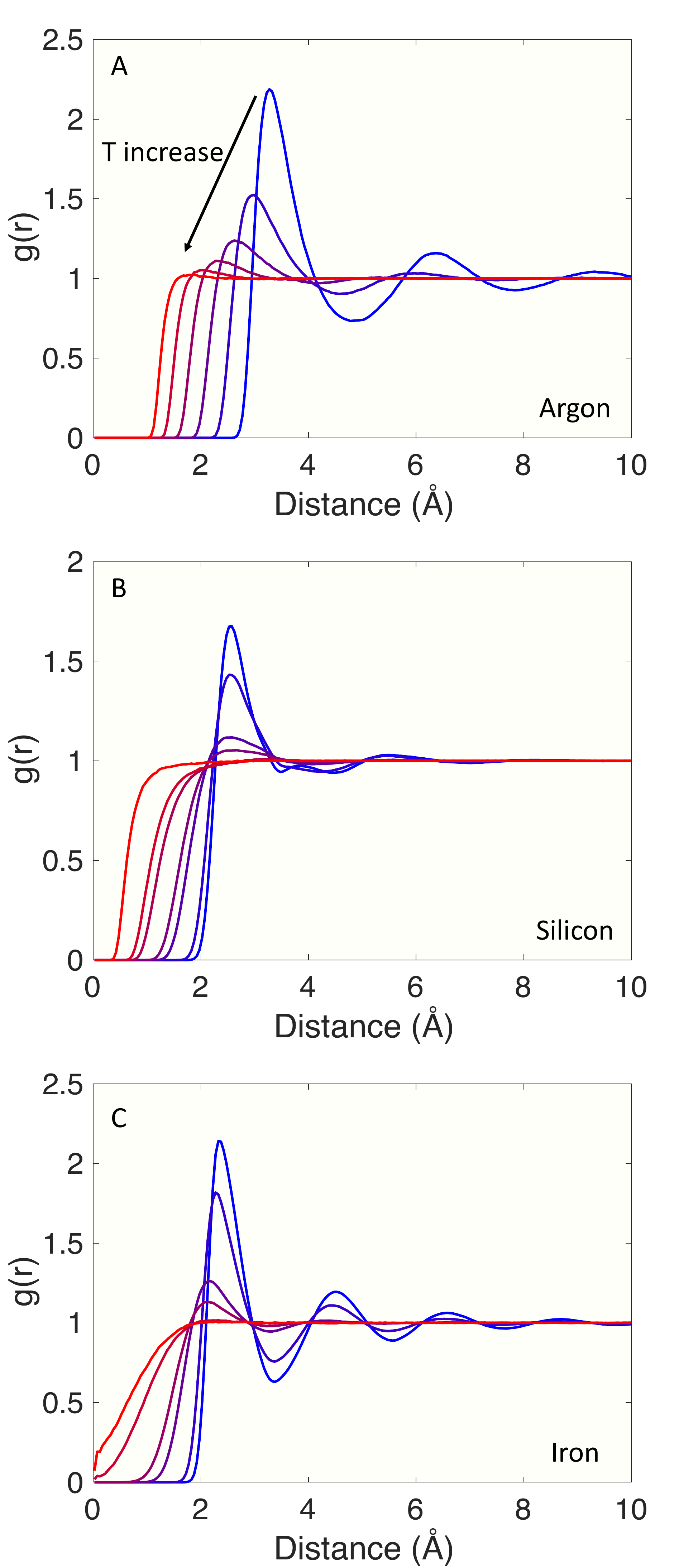}
	\caption{Pair distribution function, $g(r)$, of (A) argon from 10\textsuperscript{3} to 10\textsuperscript{8} K, (B) silicon from $5 \times 10^3$ to 10\textsuperscript{7} K, and (C) iron from $5 \times 10^3$ to 10\textsuperscript{6} K. All systems were evaluated at $\rho_0$. Lightest blue and darkest red curves represent lowest and highest temperatures, respectively. Well-defined peaks and valleys in $g(r)$ progressively disappear as temperature increases highlighting transition from liquid states at lower T to gas states at higher T. }
	\label{fig:PDF}
\end{figure}

To demonstrate that we are sampling different liquid and gas states, we examine pair distribution functions (PDF), $g(r)$, of all $\rho_0$ systems.  PDF is defined as $g(r)=\frac{1}{4\pi N n r^2}\sum_{i,j} \langle \delta(r- |\boldsymbol{r}_i - \boldsymbol{r}_j|)\rangle$ where $N$ is the number of atoms, $n$ is the number density, $\boldsymbol{r}_i$ is the atomic position of the $i$th atom, and the angled bracket denotes an ensemble average. PDFs above melting temperatures are shown in Fig. \ref{fig:PDF}. As we increase the temperature, we observe progressive disappearance of well-defined peaks and valleys in the $g(r)$ highlighting a transition from liquid to dense gas states. The first distance point at which $g(r)$ becomes finite also decreases with increase in temperature as the distance where potential energy $\sim k_BT$ becomes smaller, as expected. Thus, the effective atomic diameter becomes smaller with increase in temperature and the effective density decreases.  The compressibility factor, $Z = \frac{pV}{Nk_BT}$, which is a measure of ideal gasness, is around 1.05 for the highest temperatures for all systems, confirming that our systems are indeed in gas states despite the high nominal density.

\begin{figure}
	\centering
	\includegraphics[width=0.48\linewidth]{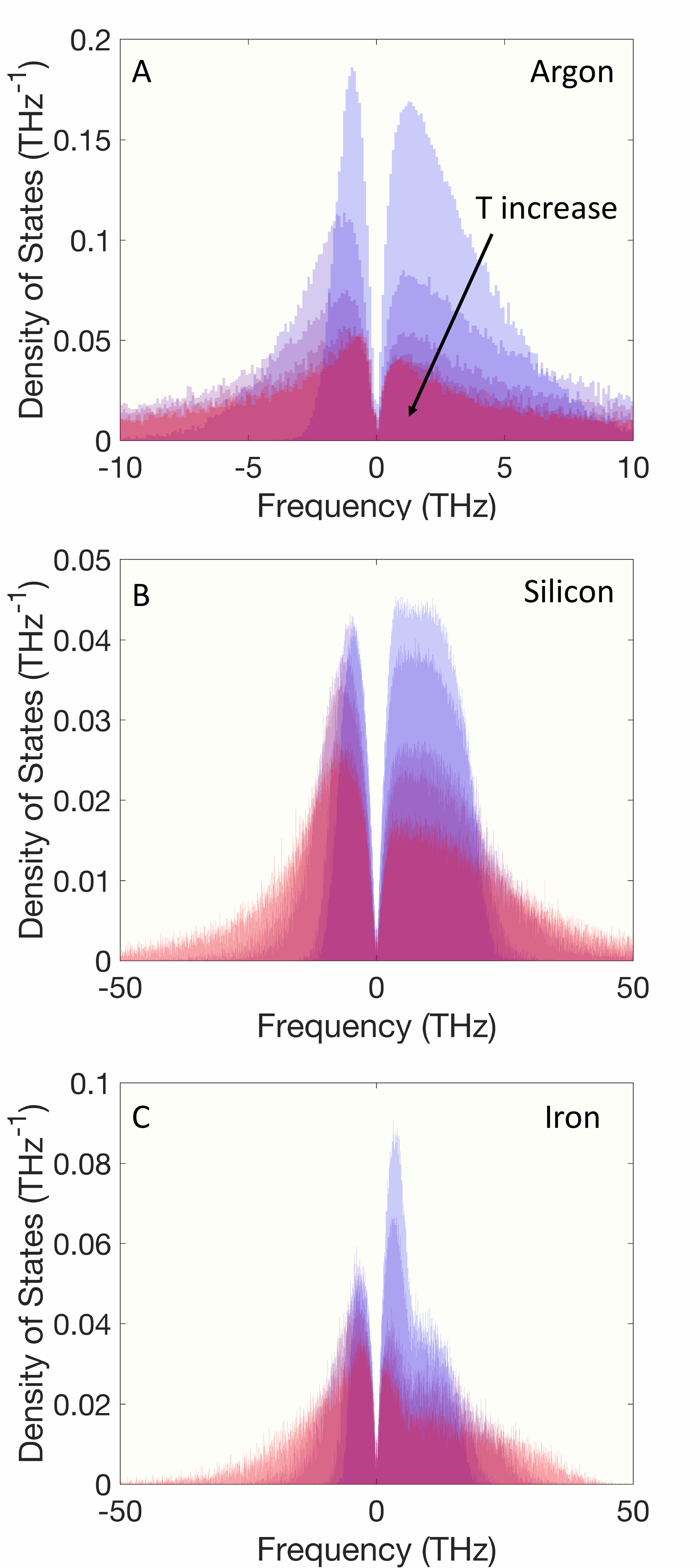}
	\caption{Instantaneous normal mode density of states (INM($\omega$)) of (A) argon from 10\textsuperscript{3} to 10\textsuperscript{8} K, (B) silicon from $5 \times 10^3$ to 10\textsuperscript{7} K, and (C) iron from $5 \times 10^3$ to 10\textsuperscript{6} K. All systems were evaluated at $\rho_0$. Each INM($\omega$) is normalized such that the integral over the entire frequency range is unity. Lightest blue and darkest red represent lowest and highest temperatures, respectively. Negative frequency in the figure means imaginary frequency. At low temperatures, modes with real frequency modes dominate the spectra. As temperature increases, however, we observe that imaginary mode populations become more significant.}
	\label{fig:DOS_LD}
\end{figure}

We now characterize their effective heat carriers. We first examine their instantaneous normal modes by diagonalizing their dynamical matrices. Resulting instantaneous normal mode densities of states, INM($\omega$), for all $\rho_0$ systems beyond the melting temperatures are shown in Fig. \ref{fig:DOS_LD}. We have done lattice dynamics calculations on small systems consisting of $\sim$ 600 atoms and larger systems of $\sim$ 8000 atoms at select temperatures for which we did not observe qualitative differences in the spectral distributions. We expect slight size effects at low frequencies $\omega  \lesssim 1$ THz; however, low frequency modes constitute only a small portion of the overall mode population and should not affect our results. In addition, specific heat does not have strong size effects when normalized by number of atoms so calculations on systems with $\sim$ 2000 atoms are sufficient here. Negative frequencies denote modes with imaginary frequencies arising from instabilities of the structure. Lightest blue shades represent lowest temperatures considered above the melting temperature and the shades become progressively red with increase in temperature. At the lowest temperatures, INM($\omega$) for all systems are dominated by real modes as measured by the areas under the curves. As temperature is increased, however, imaginary mode populations become more prominent as observed previously \cite{keyes_instantaneous_1997}. 

We note that relaxed solids have only real modes and we observe that numbers of real and imaginary modes become nearly equal at high temperatures in the gas limit from our calculations. The square of the instantaneous normal mode frequency reflects the curvature of the local potential energy landscape (PEL) that atoms participating in the normal mode see at that instant. The concept of the PEL \cite{debenedetti_supercooled_2001} is usually applied to the whole system, but here we expand the concept and define the PEL for the subsystem of the normal mode. Such a limited view of the PEL is not new. For instance, when a relaxation of glass is considered, only the atoms that are involved in the relaxation phenomenon are taken into account in depicting the PEL \cite{goldstein_viscous_1969, debenedetti_supercooled_2001}. At low temperatures the system is largely trapped in the valleys of the PEL, so the instantaneous normal mode frequencies are mostly real. In contrast at very high temperatures the system samples the positions with positive curvature (valleys) as well as the positions with negative curvature (hills) equally. For symmetric sparse random matrices, the eigenvalue distribution results in Wigner's semi-circle law with even number of positive and negative eigenvalues \cite{wigner_distribution_1958}. Thus, it is reasonable that the number of positive and negative eigenvalues are equal for dynamical matrices of high temperature gases. To describe this transition of instantaneous normal mode spectra from solid to gas, we propose two phenomenological parameters called instability parameters (IP\textsubscript{1} and IP\textsubscript{2}) as a measure of how unstable the system is in the configurational space. We define $\text{IP}_1 = \frac{2N_i}{3N}$, where $N_i$ is the total number of imaginary modes in our quantized systems, as a measure of the `gasness' of the system. IP\textsubscript{1} is linearly related to the fraction of the PEL with negative curvature that the system sees. We also define $\text{IP}_2 = \frac{N_i}{3N-N_i}$, which represents the ratio of the portion of the PELs with negative curvature to those with positive curvature. In both cases, IP\textsubscript{1,2} $= 0$ represents solid, IP\textsubscript{1,2} $= 1$ represents gas, and liquids are described by IP\textsubscript{1,2} values in-between.  Temperature dependent IP\textsubscript{1,2} values for all $\rho_0$ systems are shown in Fig. S2 in Supplementary Materials. IP\textsubscript{1,2} increases with temperature for all systems as expected and continues to increase slightly in the gas phase, possibly due to finite size effects.

\begin{figure}[h!]
	\centering
	\includegraphics[width=1\linewidth]{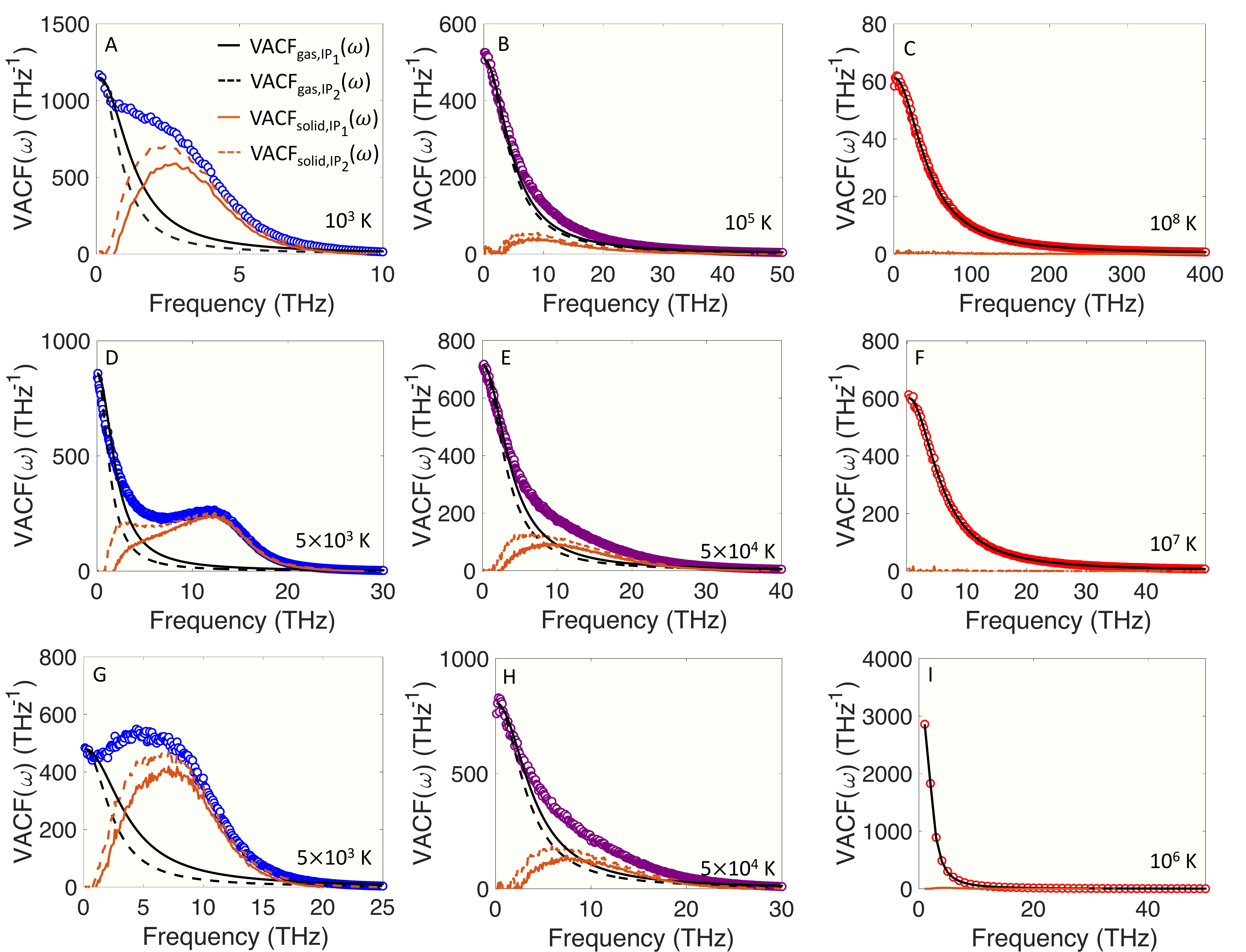}
	\caption{Decomposition of spectral velocity autocorrelation function, VACF($\omega$) at select temperatures of (A-C) argon  from 10\textsuperscript{3} to 10\textsuperscript{8} K, (D-F) silicon from $5 \times 10^3$ to 10\textsuperscript{7} K, and (G-I) iron from $5 \times 10^3$ to 10\textsuperscript{6} K. All systems were evaluated at $\rho_0$. Solid and dashed curves represent results based on IP\textsubscript{1} and IP\textsubscript{2}, respectively. Black curves represent gas-like degrees of freedom spectra, VACF\textsubscript{real}($\omega$), from instability parameter and orange curves represent solid-like degrees of freedom spectra, VACF\textsubscript{phonon}($\omega$), obtained from subtraction of VACF($\omega$) by the black curves.}
	\label{fig:DOS_VACF}
\end{figure}

With these observations in INM($\omega$), we turn to velocity autocorrelation spectra (calculated via MD simulations) of liquids and gases and see if we can obtain a consistent picture of effective heat carriers existing in solid, liquid, and gas phases. The spectral velocity autocorrelation function, VACF($\omega$), is a useful tool to study atomic dynamics. VACF($\omega$) describes phonon density of states for solids under harmonic approximations and are equivalent to INM($\omega$)  \cite{dove_introduction_1993} as demonstrated in Fig. S3 in Supplementary Materials for crystalline silicon at 1 K. VACF($\omega$) can also describe non-phononic dynamics as VACF($\omega$ = 0) represents the self-diffusion coefficient in the system (in our normalization, $D = \frac{k_BT}{12mN}\text{VACF}(\omega = 0)$ where $D$ and $m$ represent self-diffusion coefficient and atomic mass, respectively). As such, there have been prior attempts to decompose VACF($\omega$) into diffusion and phonon contributions from both MD simulations and experiments utilizing Langevin-Brownian motion and hard-sphere diffusion processes with varying degrees of success \cite{rahman_correlations_1964, verkerk_velocity_1989}. We take a different approach and examine INM($\omega$) and VACF($\omega$) together to characterize the heat carriers. At high temperatures, we observe Lorentzian lineshapes for all gas phases demonstrated in Fig. \ref{fig:DOS_VACF} (C, I, F) as expected. Similar to INM($\omega$), integration of VACF($\omega$) over frequency leads to the total translational atomic degrees of freedom, $3N$. This property is where we make connections between VACF($\omega$) and INM($\omega$). In the case of a gas (IP\textsubscript{1,2} $= 1$ for our analysis), VACF($\omega$) is Lorentzian with a peak given by $D$ and integral describing all $3N$ atomic degrees of freedom, no phonon quasi-particles. For a solid (IP\textsubscript{1,2} $= 0$), VACF($\omega$) strictly describes phonon density of states, with no free diffusing atoms. Motivated by this contrast, we partition VACF($\omega$) into a gas-like portion, $\text{VACF}_{\text{gas}}(\omega)$, with negligible attractive potential interactions and a solid-like portion, $\text{VACF}_{\text{solid}}(\omega)$, with strong potential interactions such that 
\begin{equation}
    \frac{1}{3N}\int \text{VACF}_{\text{gas}}(\omega) d\omega = \text{IP}_{1,2} 
\end{equation}
\begin{equation}
    \frac{1}{3N}\int \text{VACF}_{\text{solid}}(\omega) d\omega = 1 - \text{IP}_{1,2} 
\end{equation}
VACF\textsubscript{gas}($\omega$) is assumed to have a Lorentzian lineshape with the height and width determined by the diffusion coefficient and IP\textsubscript{1,2}, respectively. VACF\textsubscript{solid}($\omega$) is then determined by subtracting VACF($\omega$) by VACF\textsubscript{gas}($\omega$). Calculated VACF($\omega$) and corresponding gas-like and solid-like decompositions for all systems at select low, intermediate, and high temperatures are shown in Fig. \ref{fig:DOS_VACF}. If the `gasness' in the system is severely overpredicted, hence affecting the Lorenzian linewidth, $\text{VACF}_{\text{solid}}(\omega)$ will be negative over a wide range of frequencies, which is unphysical. The results here satisfy this test. 

We first note that solid-like populations continuously decrease with increasing temperature as expected from other prior phonon works \cite{bolmatov_phonon_2012,zaccone_universal_2021, baggioli_explaining_2021} on liquids. However, rather than phononic degrees of freedom simply disappearing \cite{zaccone_universal_2021, baggioli_explaining_2021}, they transition to gas-like degrees of freedom and are eventually all absorbed into the gas-like population at high temperatures as shown in Fig. \ref{fig:DOS_VACF} (C, F, I). This observation in combination with our pair distribution function analysis at high temperatures (see Fig. \ref{fig:PDF}) leads to an important insight into INM($\omega$) of liquid and gas states: real frequency modes do not have the same meaning as conventional harmonic oscillators (phonons) in solids. Prior instantaneous normal mode works have generally focused on the origin of imaginary modes alone and simply considered real modes to derive from harmonic oscillators in solids \cite{stassen_instantaneous_1994, melzer_instantaneous_2012}. However, average collision frequencies for high temperature liquids and gases, conservatively estimated by inverse of the time required for atoms to travel one interatomic distance ($(V/N)^{1/3}$) is well above 100 THz, higher than most of the real normal mode frequencies in Fig. \ref{fig:DOS_LD}. There are nearly $\frac{3}{2}N$ modes with real frequencies in INM($\omega$) for high temperature gases and interpreting them as having conventional phonon harmonic oscillators is highly questionable. Our work connecting INM($\omega$) and VACF($\omega$) calls for a new interpretation of these real frequency modes for liquid and gas states in terms of the local curvature in the PEL, and using the instability parameter as a measure of `gasness' to interpret the correlated INM and VACF behaviors. 

With VACF($\omega$) decomposed into VACF\textsubscript{solid}($\omega$) and VACF\textsubscript{gas}($\omega$), we now assess heat capacity contributions from the separate solid-like and gas-like pictures. We assume that quadratic terms in the time-dependent potential energy Taylor expansion is dominant for the constant volume systems with $\rho_0$ for VACF\textsubscript{solid}($\omega$). The total energy considered here under classical approximation is given by 
\begin{equation}
    E = \int d\omega  \text{VACF}_{solid}(\omega) k_BT + \int d\omega \text{VACF}_{gas}(\omega) \frac{1}{2}k_BT 
    \label{eq:E}
\end{equation}
Eq. \ref{eq:E} can be further simplified in terms of IP\textsubscript{1,2} by $E = (1-\text{IP}_{1,2}) (3Nk_BT) + \text{IP}_{1,2}\big(\frac{3}{2}Nk_BT)$. The corresponding constant volume heat capacity for solids, liquids, and gases under the classical and harmonic approximations is then simply
\begin{equation}
C_{V, \text{IP}_{1,2}} = (1-\text{IP}_{1,2})(3Nk_B) +  \text{IP}_{1,2}\Big(\frac{3}{2}Nk_B \Big) - \frac{d\text{IP}_{1,2}}{dT}\Big(\frac{3}{2}Nk_BT\Big)
\label{eq:cv}
\end{equation}
Instability parameter derivatives at each $T$ were found from additional instability parameters calculated at adjacent temperatures ($0.9T$ and $1.1T$), i.e. numerical derivatives. 

\begin{figure}[h!]
	\centering
	\includegraphics[width=0.67\linewidth]{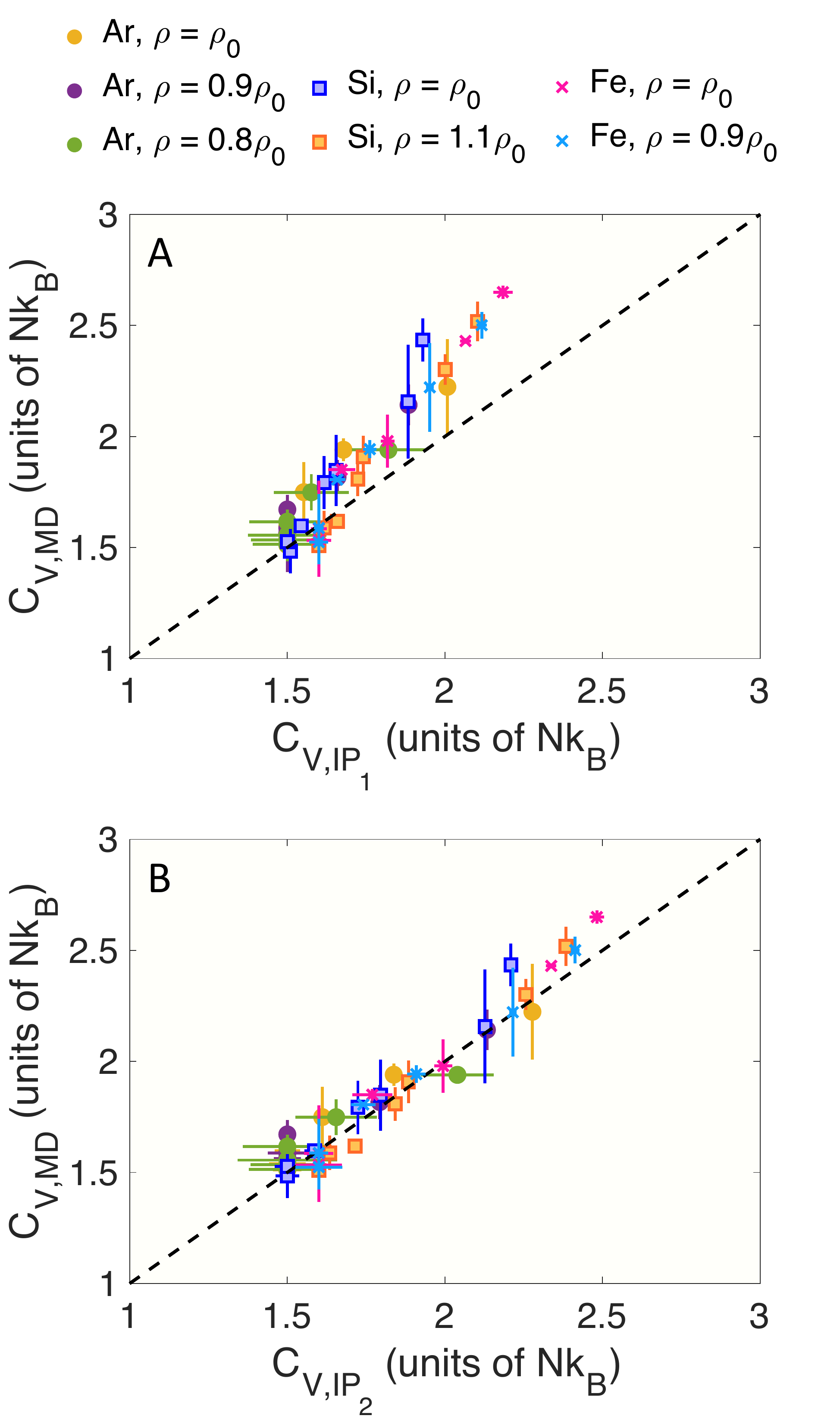}
	\caption{Predicted constant volume heat capacity of various monatomic systems with different densities from instability parameters (Eq. \ref{eq:cv}), compared to independently calculated heat capacity from molecular dynamics. Dashed black lines represent one to one correspondence. For $C_{V, \text{IP}_2}$, deviations are within 5\%.  }
	\label{fig:Cv_f}
\end{figure}

We can now compare our predictions from Eq. \ref{eq:cv} with heat capacity independently calculated by $C_{V, MD} = \frac{\langle E \rangle^2 - \langle E^2 \rangle}{k_BT^2}$ from molecular dynamics (see Fig. \ref{fig:Cv_f}). We see good agreement between our predictions from Eq. \ref{eq:cv} and MD specific heat values for all systems. Clear convergence to the gas limit of $C_V = 1.5Nk_B$ at IP $= 1$ is also observed as predicted (see Fig. S5 in Supplementary Materials). $C_{V, \text{IP}_{1}}$ tends to underestimate $C_{V,MD}$ slightly, whereas $C_{V, \text{IP}_{2}}$ shows a better overall agreement. However, in spite of small differences, these observations give strong evidence supporting our interpretations of effective heat carriers in INM($\omega$) and VACF($\omega$) of solid, liquid, and gas phases from a unified perspective via simply defined microscopic instability parameters. 

We anticipate two possible sources of error when comparing the specific heat values. The first one is the theory itself, i.e., IP\textsubscript{1,2} does not represent `gasness' correctly. The other source of error is neglecting anharmonicity in potential interactions. To minimize the effect of volume expansion/contraction on the potential interactions, we have focused our discussion primarily to $\rho_0$ systems. Temperature dependent MD specific heat values are shown in Fig. S4 in Supplementary Materials. We have also computed specific heat and IP\textsubscript{1,2} for densities slightly different from $\rho_0$ to examine the sensitivity of volume effect near $\rho_0$ as shown in Fig. \ref{fig:Cv_f}. We note that our work does not assume Debye or Gaussian densities of states that are often used in the literature \cite{bolmatov_phonon_2012, zaccone_universal_2021, baggioli_explaining_2021}. Rather, we examined the actual INM($\omega$) and VACF($\omega$) from realistic potentials and make connections between the two spectra of liquid and gas systems. Direct comparison with experiments is challenging as most heat capacity measurements are done under constant pressure conditions. Further, it is expected that anharmonicity will be important near melting and glass transition temperatures. Pressure and volume dependence and the role of anharmonicity in heat capacity merit further investigations.
 
We expect that our work will also be useful in studying thermodynamics and thermal properties of non-conventional materials including liquid crystals and solid ionic conductors. There has been a lot of recent research interests for thermoelectric power generators in certain solid ionic conductors dubbed `phonon-liquid, electron crystal' where atoms at sub-lattice sites are fixed while others are diffusing \cite{liu_copper_2012, voneshen_hopping_2017, niedziela_selective_2019}. This leads to desirable low heat capacity and non-electronic thermal conductivity but the origin of these is not clear. It is possible that our work helps identify the mechanism behind these complex phenomena.

In summary, we have addressed thermodynamics of liquids from both solid and gas perspectives. We propose to characterize effective heat carriers in liquids via instability parameters describing `gasness' in both instantaneous normal mode and velocity autocorrelation spectra. In our approach we interpret instantaneous normal modes reflecting the local curvature of the potential energy landscape, rather than considering the instantaneous normal modes with real frequencies equivalent to harmonic oscillators in solids as is often done in the literature. We provide strong evidence in support of our proposal by good agreement between predicted specific heat and specific heat values calculated from molecular dynamics. Our work provides some insights into the long-standing problem of thermodynamics of liquids and suggests pathways to a unified framework in studying thermodynamics of solid, liquid, and gas phases. 

This research was supported by the U.S. Department of Energy, Office of Science, Basic Energy Sciences, Materials Sciences and Engineering Division. This work used the Extreme Science and
Engineering Discovery Environment (XSEDE) Expanse under
Allocation No. TG-MAT200012. This research used resources of the National Energy Research Scientific Computing Center (NERSC), a U.S. Department of Energy Office of Science User Facility located at Lawrence Berkeley National Laboratory, operated under Contract No. DE-AC02-05CH11231 using NERSC award BES-ERCAPERCAP0020503. 

\clearpage
\clearpage
\clearpage
\clearpage
\clearpage


\begin{thebibliography}{56}%
\makeatletter
\providecommand \@ifxundefined [1]{%
 \@ifx{#1\undefined}
}%
\providecommand \@ifnum [1]{%
 \ifnum #1\expandafter \@firstoftwo
 \else \expandafter \@secondoftwo
 \fi
}%
\providecommand \@ifx [1]{%
 \ifx #1\expandafter \@firstoftwo
 \else \expandafter \@secondoftwo
 \fi
}%
\providecommand \natexlab [1]{#1}%
\providecommand \enquote  [1]{``#1''}%
\providecommand \bibnamefont  [1]{#1}%
\providecommand \bibfnamefont [1]{#1}%
\providecommand \citenamefont [1]{#1}%
\providecommand \href@noop [0]{\@secondoftwo}%
\providecommand \href [0]{\begingroup \@sanitize@url \@href}%
\providecommand \@href[1]{\@@startlink{#1}\@@href}%
\providecommand \@@href[1]{\endgroup#1\@@endlink}%
\providecommand \@sanitize@url [0]{\catcode `\\12\catcode `\$12\catcode
  `\&12\catcode `\#12\catcode `\^12\catcode `\_12\catcode `\%12\relax}%
\providecommand \@@startlink[1]{}%
\providecommand \@@endlink[0]{}%
\providecommand \url  [0]{\begingroup\@sanitize@url \@url }%
\providecommand \@url [1]{\endgroup\@href {#1}{\urlprefix }}%
\providecommand \urlprefix  [0]{URL }%
\providecommand \Eprint [0]{\href }%
\providecommand \doibase [0]{https://doi.org/}%
\providecommand \selectlanguage [0]{\@gobble}%
\providecommand \bibinfo  [0]{\@secondoftwo}%
\providecommand \bibfield  [0]{\@secondoftwo}%
\providecommand \translation [1]{[#1]}%
\providecommand \BibitemOpen [0]{}%
\providecommand \bibitemStop [0]{}%
\providecommand \bibitemNoStop [0]{.\EOS\space}%
\providecommand \EOS [0]{\spacefactor3000\relax}%
\providecommand \BibitemShut  [1]{\csname bibitem#1\endcsname}%
\let\auto@bib@innerbib\@empty
\bibitem [{\citenamefont {Williams}\ \emph {et~al.}(2006)\citenamefont
  {Williams}, \citenamefont {Toth},\ and\ \citenamefont
  {Clarno}}]{williams_assessment_2006-1}%
  \BibitemOpen
  \bibfield  {author} {\bibinfo {author} {\bibfnamefont {D.~F.}\ \bibnamefont
  {Williams}}, \bibinfo {author} {\bibfnamefont {L.~M.}\ \bibnamefont {Toth}},\
  and\ \bibinfo {author} {\bibfnamefont {K.~T.}\ \bibnamefont {Clarno}},\
  }\href@noop {} {\emph {\bibinfo {title} {Assessment of {Candidate} {Molten}
  {Salt} {Coolants} for the {Advanced} {High} {Temperature} {Reactor}
  ({AHTR}).}}}\ (\bibinfo  {publisher} {United States. Department of Energy},\
  \bibinfo {year} {2006})\BibitemShut {NoStop}%
\bibitem [{\citenamefont {Li}(2016)}]{li_sensible_2016}%
  \BibitemOpen
  \bibfield  {author} {\bibinfo {author} {\bibfnamefont {G.}~\bibnamefont
  {Li}},\ }\bibfield  {title} {\bibinfo {title} {Sensible heat thermal storage
  energy and exergy performance evaluations},\ }\href
  {https://doi.org/10.1016/j.rser.2015.09.006} {\bibfield  {journal} {\bibinfo
  {journal} {Renewable and Sustainable Energy Reviews}\ }\textbf {\bibinfo
  {volume} {53}},\ \bibinfo {pages} {897} (\bibinfo {year} {2016})}\BibitemShut
  {NoStop}%
\bibitem [{\citenamefont {de~La~Tour}(1822)}]{de_la_tour_expose_1822}%
  \BibitemOpen
  \bibfield  {author} {\bibinfo {author} {\bibfnamefont {C.~C.}\ \bibnamefont
  {de~La~Tour}},\ }\bibfield  {title} {\bibinfo {title} {Exposé de quelques
  résultats obtenu par l’action combinée de la chaleur et de la compression
  sur certains liquides, tels que l’eau, l’alcool, l’éther sulfurique et
  l’essence de pétrole rectifiée},\ }\href@noop {} {\bibfield  {journal}
  {\bibinfo  {journal} {Ann. Chim. Phys}\ }\textbf {\bibinfo {volume} {21}},\
  \bibinfo {pages} {127} (\bibinfo {year} {1822})}\BibitemShut {NoStop}%
\bibitem [{\citenamefont {Faraday}(1823)}]{faraday_xvii_1823}%
  \BibitemOpen
  \bibfield  {author} {\bibinfo {author} {\bibfnamefont {M.}~\bibnamefont
  {Faraday}},\ }\bibfield  {title} {\bibinfo {title} {{XVII}. {On} the
  condensation of several gases into liquids},\ }\href
  {https://doi.org/10.1098/rstl.1823.0019} {\bibfield  {journal} {\bibinfo
  {journal} {Philosophical Transactions of the Royal Society of London}\
  }\textbf {\bibinfo {volume} {113}},\ \bibinfo {pages} {189} (\bibinfo {year}
  {1823})}\BibitemShut {NoStop}%
\bibitem [{\citenamefont {Faraday}(1971)}]{faraday_selected_1971}%
  \BibitemOpen
  \bibfield  {author} {\bibinfo {author} {\bibfnamefont {M.}~\bibnamefont
  {Faraday}},\ }\href@noop {} {\emph {\bibinfo {title} {The {Selected}
  {Correspondence} of {Michael} {Faraday}: 1812-1848}}},\ Vol.~\bibinfo
  {volume} {1}\ (\bibinfo  {publisher} {Cambridge University Press},\ \bibinfo
  {year} {1971})\BibitemShut {NoStop}%
\bibitem [{\citenamefont {Mendelejeff}(1861)}]{mendelejeff_ueber_1861}%
  \BibitemOpen
  \bibfield  {author} {\bibinfo {author} {\bibfnamefont {D.}~\bibnamefont
  {Mendelejeff}},\ }\bibfield  {title} {\bibinfo {title} {Ueber die
  {Ausdehnung} der {Flüssigkeiten} beim {Erwärmen} über ihren
  {Siedepunkt}},\ }\href {https://doi.org/10.1002/jlac.18611190102} {\bibfield
  {journal} {\bibinfo  {journal} {Justus Liebigs Annalen der Chemie}\ }\textbf
  {\bibinfo {volume} {119}},\ \bibinfo {pages} {1} (\bibinfo {year}
  {1861})}\BibitemShut {NoStop}%
\bibitem [{\citenamefont {Andrews}(1869)}]{andrews_bakerian_1869}%
  \BibitemOpen
  \bibfield  {author} {\bibinfo {author} {\bibfnamefont {T.}~\bibnamefont
  {Andrews}},\ }\bibfield  {title} {\bibinfo {title} {The {Bakerian} {Lecture}:
  {On} the {Continuity} of the {Gaseous} and {Liquid} {States} of {Matter}},\
  }\href {https://www.jstor.org/stable/109009} {\bibfield  {journal} {\bibinfo
  {journal} {Philosophical Transactions of the Royal Society of London}\
  }\textbf {\bibinfo {volume} {159}},\ \bibinfo {pages} {575} (\bibinfo {year}
  {1869})}\BibitemShut {NoStop}%
\bibitem [{\citenamefont {Tait}\ and\ \citenamefont
  {Brown}(1889)}]{tait_scientific_1889}%
  \BibitemOpen
  \bibfield  {author} {\bibinfo {author} {\bibfnamefont {P.~G.}\ \bibnamefont
  {Tait}}\ and\ \bibinfo {author} {\bibfnamefont {A.~C.}\ \bibnamefont
  {Brown}},\ }\href@noop {} {\emph {\bibinfo {title} {The {Scientific} {Papers}
  of the {Late} {Thomas} {Andrews}, {MD}, {FRS}, {Vice} {President} and
  {Professor} of {Chemistry}, {Queen}'s {College}, {Belfast}}}}\ (\bibinfo
  {publisher} {Macmillanand Company},\ \bibinfo {year} {1889})\BibitemShut
  {NoStop}%
\bibitem [{\citenamefont {van~der Waals}(1873)}]{van_der_waals_over_1873}%
  \BibitemOpen
  \bibfield  {author} {\bibinfo {author} {\bibfnamefont {J.~D.}\ \bibnamefont
  {van~der Waals}},\ }\href@noop {} {\emph {\bibinfo {title} {Over de
  {Continuiteit} van den {Gas}-en {Vloeistoftoestand}}}},\ Vol.~\bibinfo
  {volume} {1}\ (\bibinfo  {publisher} {Sijthoff},\ \bibinfo {year}
  {1873})\BibitemShut {NoStop}%
\bibitem [{\citenamefont {Chandler}\ \emph {et~al.}(1983)\citenamefont
  {Chandler}, \citenamefont {Weeks},\ and\ \citenamefont
  {Andersen}}]{chandler_van_1983}%
  \BibitemOpen
  \bibfield  {author} {\bibinfo {author} {\bibfnamefont {D.}~\bibnamefont
  {Chandler}}, \bibinfo {author} {\bibfnamefont {J.~D.}\ \bibnamefont
  {Weeks}},\ and\ \bibinfo {author} {\bibfnamefont {H.~C.}\ \bibnamefont
  {Andersen}},\ }\bibfield  {title} {\bibinfo {title} {Van {Der} {Waals}
  {Picture} of {Liquids}, {Solids}, and {Phase} {Transformations}},\ }\href
  {https://www.jstor.org/stable/1690308} {\bibfield  {journal} {\bibinfo
  {journal} {Science}\ }\textbf {\bibinfo {volume} {220}},\ \bibinfo {pages}
  {787} (\bibinfo {year} {1983})}\BibitemShut {NoStop}%
\bibitem [{\citenamefont {Dyre}(2016)}]{dyre_simple_2016}%
  \BibitemOpen
  \bibfield  {author} {\bibinfo {author} {\bibfnamefont {J.~C.}\ \bibnamefont
  {Dyre}},\ }\bibfield  {title} {\bibinfo {title} {Simple liquids’
  quasiuniversality and the hard-sphere paradigm},\ }\href
  {https://doi.org/10.1088/0953-8984/28/32/323001} {\bibfield  {journal}
  {\bibinfo  {journal} {Journal of Physics: Condensed Matter}\ }\textbf
  {\bibinfo {volume} {28}},\ \bibinfo {pages} {323001} (\bibinfo {year}
  {2016})}\BibitemShut {NoStop}%
\bibitem [{\citenamefont {Mie}(1903)}]{mie_zur_1903}%
  \BibitemOpen
  \bibfield  {author} {\bibinfo {author} {\bibfnamefont {G.}~\bibnamefont
  {Mie}},\ }\bibfield  {title} {\bibinfo {title} {Zur kinetischen {Theorie} der
  einatomigen {Körper}},\ }\href {https://doi.org/10.1002/andp.19033160802}
  {\bibfield  {journal} {\bibinfo  {journal} {Annalen der Physik}\ }\textbf
  {\bibinfo {volume} {316}},\ \bibinfo {pages} {657} (\bibinfo {year}
  {1903})}\BibitemShut {NoStop}%
\bibitem [{\citenamefont {Mott}(1934)}]{mott_resistance_1934}%
  \BibitemOpen
  \bibfield  {author} {\bibinfo {author} {\bibfnamefont {N.~F.}\ \bibnamefont
  {Mott}},\ }\bibfield  {title} {\bibinfo {title} {The resistance of liquid
  metals},\ }\href {https://doi.org/10.1098/rspa.1934.0166} {\bibfield
  {journal} {\bibinfo  {journal} {Proceedings of the Royal Society of London.
  Series A, Containing Papers of a Mathematical and Physical Character}\
  }\textbf {\bibinfo {volume} {146}},\ \bibinfo {pages} {465} (\bibinfo {year}
  {1934})}\BibitemShut {NoStop}%
\bibitem [{\citenamefont {Lennard-Jones}\ and\ \citenamefont
  {Devonshire}(1937)}]{lennard-jones_critical_1937}%
  \BibitemOpen
  \bibfield  {author} {\bibinfo {author} {\bibfnamefont {J.~E.}\ \bibnamefont
  {Lennard-Jones}}\ and\ \bibinfo {author} {\bibfnamefont {A.~F.}\ \bibnamefont
  {Devonshire}},\ }\bibfield  {title} {\bibinfo {title} {Critical phenomena in
  gases - {I}},\ }\href {https://doi.org/10.1098/rspa.1937.0210} {\bibfield
  {journal} {\bibinfo  {journal} {Proceedings of the Royal Society of London.
  Series A - Mathematical and Physical Sciences}\ }\textbf {\bibinfo {volume}
  {163}},\ \bibinfo {pages} {53} (\bibinfo {year} {1937})}\BibitemShut
  {NoStop}%
\bibitem [{\citenamefont {Granato}(2002)}]{granato_specific_2002}%
  \BibitemOpen
  \bibfield  {author} {\bibinfo {author} {\bibfnamefont {A.}~\bibnamefont
  {Granato}},\ }\bibfield  {title} {\bibinfo {title} {The specific heat of
  simple liquids},\ }\href {https://doi.org/10.1016/S0022-3093(02)01498-9}
  {\bibfield  {journal} {\bibinfo  {journal} {Journal of Non-Crystalline
  Solids}\ }\textbf {\bibinfo {volume} {307-310}},\ \bibinfo {pages} {376}
  (\bibinfo {year} {2002})}\BibitemShut {NoStop}%
\bibitem [{\citenamefont {Frenkel}(1926)}]{frenkel_uber_1926}%
  \BibitemOpen
  \bibfield  {author} {\bibinfo {author} {\bibfnamefont {J.}~\bibnamefont
  {Frenkel}},\ }\bibfield  {title} {\bibinfo {title} {Über die
  {Wärmebewegung} in festen und flüssigen {Körpern}},\ }\href
  {https://doi.org/10.1007/BF01379812} {\bibfield  {journal} {\bibinfo
  {journal} {Zeitschrift für Physik}\ }\textbf {\bibinfo {volume} {35}},\
  \bibinfo {pages} {652} (\bibinfo {year} {1926})}\BibitemShut {NoStop}%
\bibitem [{\citenamefont {Frenkel}(1935)}]{frenkel_continuity_1935}%
  \BibitemOpen
  \bibfield  {author} {\bibinfo {author} {\bibfnamefont {J.}~\bibnamefont
  {Frenkel}},\ }\bibfield  {title} {\bibinfo {title} {Continuity of the {Solid}
  and the {Liquid} {States}},\ }\href {https://doi.org/10.1038/136167a0}
  {\bibfield  {journal} {\bibinfo  {journal} {Nature}\ }\textbf {\bibinfo
  {volume} {136}},\ \bibinfo {pages} {167} (\bibinfo {year}
  {1935})}\BibitemShut {NoStop}%
\bibitem [{\citenamefont {Frenkel}(1937)}]{frenkel_liquid_1937}%
  \BibitemOpen
  \bibfield  {author} {\bibinfo {author} {\bibfnamefont {J.}~\bibnamefont
  {Frenkel}},\ }\bibfield  {title} {\bibinfo {title} {On the liquid state and
  the theory of fusion},\ }\href {https://doi.org/10.1039/tf9373300058}
  {\bibfield  {journal} {\bibinfo  {journal} {Transactions of the Faraday
  Society}\ }\textbf {\bibinfo {volume} {33}},\ \bibinfo {pages} {58} (\bibinfo
  {year} {1937})}\BibitemShut {NoStop}%
\bibitem [{\citenamefont {Frenkel}(1947)}]{frenkel_kinetic_1947}%
  \BibitemOpen
  \bibfield  {author} {\bibinfo {author} {\bibfnamefont {J.}~\bibnamefont
  {Frenkel}},\ }\href@noop {} {\emph {\bibinfo {title} {Kinetic {Theory} of
  {Liquids}}}}\ (\bibinfo  {publisher} {Oxford University Press},\ \bibinfo
  {year} {1947})\BibitemShut {NoStop}%
\bibitem [{\citenamefont {Trachenko}\ and\ \citenamefont
  {Brazhkin}(2011)}]{trachenko_heat_2011}%
  \BibitemOpen
  \bibfield  {author} {\bibinfo {author} {\bibfnamefont {K.}~\bibnamefont
  {Trachenko}}\ and\ \bibinfo {author} {\bibfnamefont {V.~V.}\ \bibnamefont
  {Brazhkin}},\ }\bibfield  {title} {\bibinfo {title} {Heat capacity at the
  glass transition},\ }\href@noop {} {\bibfield  {journal} {\bibinfo  {journal}
  {Physical Review B}\ ,\ \bibinfo {pages} {6}} (\bibinfo {year}
  {2011})}\BibitemShut {NoStop}%
\bibitem [{\citenamefont {Bolmatov}\ \emph {et~al.}(2012)\citenamefont
  {Bolmatov}, \citenamefont {Brazhkin},\ and\ \citenamefont
  {Trachenko}}]{bolmatov_phonon_2012}%
  \BibitemOpen
  \bibfield  {author} {\bibinfo {author} {\bibfnamefont {D.}~\bibnamefont
  {Bolmatov}}, \bibinfo {author} {\bibfnamefont {V.~V.}\ \bibnamefont
  {Brazhkin}},\ and\ \bibinfo {author} {\bibfnamefont {K.}~\bibnamefont
  {Trachenko}},\ }\bibfield  {title} {\bibinfo {title} {The phonon theory of
  liquid thermodynamics},\ }\href {https://doi.org/10.1038/srep00421}
  {\bibfield  {journal} {\bibinfo  {journal} {Scientific Reports}\ }\textbf
  {\bibinfo {volume} {2}},\ \bibinfo {pages} {421} (\bibinfo {year}
  {2012})}\BibitemShut {NoStop}%
\bibitem [{\citenamefont {Andritsos}\ \emph {et~al.}(2013)\citenamefont
  {Andritsos}, \citenamefont {Zarkadoula}, \citenamefont {Phillips},
  \citenamefont {Dove}, \citenamefont {Walker}, \citenamefont {Brazhkin},\ and\
  \citenamefont {Trachenko}}]{andritsos_heat_2013}%
  \BibitemOpen
  \bibfield  {author} {\bibinfo {author} {\bibfnamefont {E.~I.}\ \bibnamefont
  {Andritsos}}, \bibinfo {author} {\bibfnamefont {E.}~\bibnamefont
  {Zarkadoula}}, \bibinfo {author} {\bibfnamefont {A.~E.}\ \bibnamefont
  {Phillips}}, \bibinfo {author} {\bibfnamefont {M.~T.}\ \bibnamefont {Dove}},
  \bibinfo {author} {\bibfnamefont {C.~J.}\ \bibnamefont {Walker}}, \bibinfo
  {author} {\bibfnamefont {V.~V.}\ \bibnamefont {Brazhkin}},\ and\ \bibinfo
  {author} {\bibfnamefont {K.}~\bibnamefont {Trachenko}},\ }\bibfield  {title}
  {\bibinfo {title} {The heat capacity of matter beyond the {Dulong}–{Petit}
  value},\ }\href {https://doi.org/10.1088/0953-8984/25/23/235401} {\bibfield
  {journal} {\bibinfo  {journal} {Journal of Physics: Condensed Matter}\
  }\textbf {\bibinfo {volume} {25}},\ \bibinfo {pages} {235401} (\bibinfo
  {year} {2013})}\BibitemShut {NoStop}%
\bibitem [{\citenamefont {Trachenko}\ and\ \citenamefont
  {Brazhkin}(2016)}]{trachenko_collective_2016}%
  \BibitemOpen
  \bibfield  {author} {\bibinfo {author} {\bibfnamefont {K.}~\bibnamefont
  {Trachenko}}\ and\ \bibinfo {author} {\bibfnamefont {V.~V.}\ \bibnamefont
  {Brazhkin}},\ }\bibfield  {title} {\bibinfo {title} {Collective modes and
  thermodynamics of the liquid state},\ }\href
  {https://doi.org/10.1088/0034-4885/79/1/016502} {\bibfield  {journal}
  {\bibinfo  {journal} {Reports on Progress in Physics}\ }\textbf {\bibinfo
  {volume} {79}},\ \bibinfo {pages} {016502} (\bibinfo {year} {2016})},\
  \bibinfo {note} {arXiv: 1512.06592}\BibitemShut {NoStop}%
\bibitem [{\citenamefont {Wang}\ \emph {et~al.}(2017)\citenamefont {Wang},
  \citenamefont {Yang}, \citenamefont {Dove}, \citenamefont {Fomin},
  \citenamefont {Brazhkin},\ and\ \citenamefont
  {Trachenko}}]{wang_direct_2017}%
  \BibitemOpen
  \bibfield  {author} {\bibinfo {author} {\bibfnamefont {L.}~\bibnamefont
  {Wang}}, \bibinfo {author} {\bibfnamefont {C.}~\bibnamefont {Yang}}, \bibinfo
  {author} {\bibfnamefont {M.~T.}\ \bibnamefont {Dove}}, \bibinfo {author}
  {\bibfnamefont {Y.~D.}\ \bibnamefont {Fomin}}, \bibinfo {author}
  {\bibfnamefont {V.~V.}\ \bibnamefont {Brazhkin}},\ and\ \bibinfo {author}
  {\bibfnamefont {K.}~\bibnamefont {Trachenko}},\ }\bibfield  {title} {\bibinfo
  {title} {Direct links between dynamical, thermodynamic, and structural
  properties of liquids: {Modeling} results},\ }\href
  {https://doi.org/10.1103/PhysRevE.95.032116} {\bibfield  {journal} {\bibinfo
  {journal} {Physical Review E}\ }\textbf {\bibinfo {volume} {95}},\ \bibinfo
  {pages} {032116} (\bibinfo {year} {2017})}\BibitemShut {NoStop}%
\bibitem [{\citenamefont {Yang}\ \emph {et~al.}(2017)\citenamefont {Yang},
  \citenamefont {Dove}, \citenamefont {Brazhkin},\ and\ \citenamefont
  {Trachenko}}]{yang_emergence_2017}%
  \BibitemOpen
  \bibfield  {author} {\bibinfo {author} {\bibfnamefont {C.}~\bibnamefont
  {Yang}}, \bibinfo {author} {\bibfnamefont {M.}~\bibnamefont {Dove}}, \bibinfo
  {author} {\bibfnamefont {V.}~\bibnamefont {Brazhkin}},\ and\ \bibinfo
  {author} {\bibfnamefont {K.}~\bibnamefont {Trachenko}},\ }\bibfield  {title}
  {\bibinfo {title} {Emergence and {Evolution} of the k {Gap} in {Spectra} of
  {Liquid} and {Supercritical} {States}},\ }\bibfield  {journal} {\bibinfo
  {journal} {Physical Review Letters}\ }\textbf {\bibinfo {volume} {118}},\
  \href {https://doi.org/10.1103/PhysRevLett.118.215502}
  {10.1103/PhysRevLett.118.215502} (\bibinfo {year} {2017})\BibitemShut
  {NoStop}%
\bibitem [{\citenamefont {Brazhkin}\ \emph {et~al.}(2018)\citenamefont
  {Brazhkin}, \citenamefont {Fomin}, \citenamefont {Ryzhov}, \citenamefont
  {Tsiok},\ and\ \citenamefont {Trachenko}}]{brazhkin_liquid-like_2018}%
  \BibitemOpen
  \bibfield  {author} {\bibinfo {author} {\bibfnamefont {V.~V.}\ \bibnamefont
  {Brazhkin}}, \bibinfo {author} {\bibfnamefont {Y.~D.}\ \bibnamefont {Fomin}},
  \bibinfo {author} {\bibfnamefont {V.~N.}\ \bibnamefont {Ryzhov}}, \bibinfo
  {author} {\bibfnamefont {E.~N.}\ \bibnamefont {Tsiok}},\ and\ \bibinfo
  {author} {\bibfnamefont {K.}~\bibnamefont {Trachenko}},\ }\bibfield  {title}
  {\bibinfo {title} {Liquid-like and gas-like features of a simple fluid: {An}
  insight from theory and simulation},\ }\href
  {https://doi.org/10.1016/j.physa.2018.06.084} {\bibfield  {journal} {\bibinfo
   {journal} {Physica A: Statistical Mechanics and its Applications}\ }\textbf
  {\bibinfo {volume} {509}},\ \bibinfo {pages} {690} (\bibinfo {year}
  {2018})}\BibitemShut {NoStop}%
\bibitem [{\citenamefont {Tomiyoshi}\ and\ \citenamefont
  {Ueda}(2019)}]{tomiyoshi_heat_2019}%
  \BibitemOpen
  \bibfield  {author} {\bibinfo {author} {\bibfnamefont {Y.}~\bibnamefont
  {Tomiyoshi}}\ and\ \bibinfo {author} {\bibfnamefont {D.}~\bibnamefont
  {Ueda}},\ }\bibfield  {title} {\bibinfo {title} {Heat capacity of simple
  liquids in light of hydrodynamics as {U}(1) gauge theory},\ }\href
  {https://doi.org/10.1103/PhysRevE.100.012103} {\bibfield  {journal} {\bibinfo
   {journal} {Physical Review E}\ }\textbf {\bibinfo {volume} {100}},\ \bibinfo
  {pages} {012103} (\bibinfo {year} {2019})}\BibitemShut {NoStop}%
\bibitem [{\citenamefont {Khusnutdinoff}\ \emph {et~al.}(2020)\citenamefont
  {Khusnutdinoff}, \citenamefont {Cockrell}, \citenamefont {Dicks},
  \citenamefont {Jensen}, \citenamefont {Le}, \citenamefont {Wang},
  \citenamefont {Dove}, \citenamefont {Mokshin}, \citenamefont {Brazhkin},\
  and\ \citenamefont {Trachenko}}]{khusnutdinoff_collective_2020}%
  \BibitemOpen
  \bibfield  {author} {\bibinfo {author} {\bibfnamefont {R.~M.}\ \bibnamefont
  {Khusnutdinoff}}, \bibinfo {author} {\bibfnamefont {C.}~\bibnamefont
  {Cockrell}}, \bibinfo {author} {\bibfnamefont {O.~A.}\ \bibnamefont {Dicks}},
  \bibinfo {author} {\bibfnamefont {A.~C.~S.}\ \bibnamefont {Jensen}}, \bibinfo
  {author} {\bibfnamefont {M.~D.}\ \bibnamefont {Le}}, \bibinfo {author}
  {\bibfnamefont {L.}~\bibnamefont {Wang}}, \bibinfo {author} {\bibfnamefont
  {M.~T.}\ \bibnamefont {Dove}}, \bibinfo {author} {\bibfnamefont {A.~V.}\
  \bibnamefont {Mokshin}}, \bibinfo {author} {\bibfnamefont {V.~V.}\
  \bibnamefont {Brazhkin}},\ and\ \bibinfo {author} {\bibfnamefont
  {K.}~\bibnamefont {Trachenko}},\ }\bibfield  {title} {\bibinfo {title}
  {Collective modes and gapped momentum states in liquid {Ga}: {Experiment},
  theory, and simulation},\ }\href
  {https://doi.org/10.1103/PhysRevB.101.214312} {\bibfield  {journal} {\bibinfo
   {journal} {Physical Review B}\ }\textbf {\bibinfo {volume} {101}},\ \bibinfo
  {pages} {214312} (\bibinfo {year} {2020})}\BibitemShut {NoStop}%
\bibitem [{\citenamefont {Kryuchkov}\ \emph {et~al.}(2020)\citenamefont
  {Kryuchkov}, \citenamefont {Mistryukova}, \citenamefont {Sapelkin},
  \citenamefont {Brazhkin},\ and\ \citenamefont
  {Yurchenko}}]{Kryuchkov_universal_2020}%
  \BibitemOpen
  \bibfield  {author} {\bibinfo {author} {\bibfnamefont {N.~P.}\ \bibnamefont
  {Kryuchkov}}, \bibinfo {author} {\bibfnamefont {L.~A.}\ \bibnamefont
  {Mistryukova}}, \bibinfo {author} {\bibfnamefont {A.~V.}\ \bibnamefont
  {Sapelkin}}, \bibinfo {author} {\bibfnamefont {V.~V.}\ \bibnamefont
  {Brazhkin}},\ and\ \bibinfo {author} {\bibfnamefont {S.~O.}\ \bibnamefont
  {Yurchenko}},\ }\bibfield  {title} {\bibinfo {title} {Universal {Effect} of
  {Excitation} {Dispersion} on the {Heat} {Capacity} and {Gapped} {States} in
  {Fluids}},\ }\bibfield  {journal} {\bibinfo  {journal} {Physical Review
  Letters}\ }\textbf {\bibinfo {volume} {125}},\ \href
  {https://doi.org/10.1103/PhysRevLett.125.125501}
  {10.1103/PhysRevLett.125.125501} (\bibinfo {year} {2020})\BibitemShut
  {NoStop}%
\bibitem [{\citenamefont {Zaccone}\ and\ \citenamefont
  {Baggioli}(2021)}]{zaccone_universal_2021}%
  \BibitemOpen
  \bibfield  {author} {\bibinfo {author} {\bibfnamefont {A.}~\bibnamefont
  {Zaccone}}\ and\ \bibinfo {author} {\bibfnamefont {M.}~\bibnamefont
  {Baggioli}},\ }\bibfield  {title} {\bibinfo {title} {Universal law for the
  vibrational density of states of liquids},\ }\href
  {https://doi.org/10.1073/pnas.2022303118} {\bibfield  {journal} {\bibinfo
  {journal} {Proceedings of the National Academy of Sciences}\ }\textbf
  {\bibinfo {volume} {118}},\ \bibinfo {pages} {e2022303118} (\bibinfo {year}
  {2021})}\BibitemShut {NoStop}%
\bibitem [{\citenamefont {Baggioli}\ and\ \citenamefont
  {Zaccone}(2021)}]{baggioli_explaining_2021}%
  \BibitemOpen
  \bibfield  {author} {\bibinfo {author} {\bibfnamefont {M.}~\bibnamefont
  {Baggioli}}\ and\ \bibinfo {author} {\bibfnamefont {A.}~\bibnamefont
  {Zaccone}},\ }\bibfield  {title} {\bibinfo {title} {Explaining the specific
  heat of liquids based on instantaneous normal modes},\ }\href
  {https://doi.org/10.1103/PhysRevE.104.014103} {\bibfield  {journal} {\bibinfo
   {journal} {Physical Review E}\ }\textbf {\bibinfo {volume} {104}},\ \bibinfo
  {pages} {014103} (\bibinfo {year} {2021})},\ \bibinfo {note} {publisher:
  American Physical Society}\BibitemShut {NoStop}%
\bibitem [{\citenamefont {Togo}\ and\ \citenamefont
  {Tanaka}(2015)}]{togo_first_2015}%
  \BibitemOpen
  \bibfield  {author} {\bibinfo {author} {\bibfnamefont {A.}~\bibnamefont
  {Togo}}\ and\ \bibinfo {author} {\bibfnamefont {I.}~\bibnamefont {Tanaka}},\
  }\bibfield  {title} {\bibinfo {title} {First principles phonon calculations
  in materials science},\ }\href
  {https://doi.org/10.1016/j.scriptamat.2015.07.021} {\bibfield  {journal}
  {\bibinfo  {journal} {Scripta Materialia}\ }\textbf {\bibinfo {volume}
  {108}},\ \bibinfo {pages} {1} (\bibinfo {year} {2015})}\BibitemShut {NoStop}%
\bibitem [{\citenamefont
  {Jones}(1924{\natexlab{a}})}]{jones_determination_1924}%
  \BibitemOpen
  \bibfield  {author} {\bibinfo {author} {\bibfnamefont {J.~E.}\ \bibnamefont
  {Jones}},\ }\bibfield  {title} {\bibinfo {title} {On the determination of
  molecular fields. —{II}. {From} the equation of state of a gas},\ }\href
  {https://doi.org/10.1098/rspa.1924.0082} {\bibfield  {journal} {\bibinfo
  {journal} {Proceedings of the Royal Society of London. Series A, Containing
  Papers of a Mathematical and Physical Character}\ }\textbf {\bibinfo {volume}
  {106}},\ \bibinfo {pages} {463} (\bibinfo {year}
  {1924}{\natexlab{a}})}\BibitemShut {NoStop}%
\bibitem [{\citenamefont
  {Jones}(1924{\natexlab{b}})}]{jones_determination_1924-1}%
  \BibitemOpen
  \bibfield  {author} {\bibinfo {author} {\bibfnamefont {J.~E.}\ \bibnamefont
  {Jones}},\ }\bibfield  {title} {\bibinfo {title} {On the determination of
  molecular fields.—{I}. {From} the variation of the viscosity of a gas with
  temperature},\ }\href {https://doi.org/10.1098/rspa.1924.0081} {\bibfield
  {journal} {\bibinfo  {journal} {Proceedings of the Royal Society of London.
  Series A, Containing Papers of a Mathematical and Physical Character}\
  }\textbf {\bibinfo {volume} {106}},\ \bibinfo {pages} {441} (\bibinfo {year}
  {1924}{\natexlab{b}})}\BibitemShut {NoStop}%
\bibitem [{\citenamefont {Rahman}\ \emph {et~al.}(1976)\citenamefont {Rahman},
  \citenamefont {Mandell},\ and\ \citenamefont
  {McTague}}]{rahman_molecular_1976}%
  \BibitemOpen
  \bibfield  {author} {\bibinfo {author} {\bibfnamefont {A.}~\bibnamefont
  {Rahman}}, \bibinfo {author} {\bibfnamefont {M.~J.}\ \bibnamefont
  {Mandell}},\ and\ \bibinfo {author} {\bibfnamefont {J.~P.}\ \bibnamefont
  {McTague}},\ }\bibfield  {title} {\bibinfo {title} {Molecular dynamics study
  of an amorphous {Lennard}‐{Jones} system at low temperature},\ }\href
  {https://doi.org/10.1063/1.432380} {\bibfield  {journal} {\bibinfo  {journal}
  {The Journal of Chemical Physics}\ }\textbf {\bibinfo {volume} {64}},\
  \bibinfo {pages} {1564} (\bibinfo {year} {1976})}\BibitemShut {NoStop}%
\bibitem [{\citenamefont {Stillinger}\ and\ \citenamefont
  {Weber}(1985)}]{stillinger_computer_1985}%
  \BibitemOpen
  \bibfield  {author} {\bibinfo {author} {\bibfnamefont {F.~H.}\ \bibnamefont
  {Stillinger}}\ and\ \bibinfo {author} {\bibfnamefont {T.~A.}\ \bibnamefont
  {Weber}},\ }\bibfield  {title} {\bibinfo {title} {Computer simulation of
  local order in condensed phases of silicon},\ }\href
  {http://journals.aps.org/prb/abstract/10.1103/PhysRevB.31.5262} {\bibfield
  {journal} {\bibinfo  {journal} {Physical review B}\ }\textbf {\bibinfo
  {volume} {31}},\ \bibinfo {pages} {5262} (\bibinfo {year}
  {1985})}\BibitemShut {NoStop}%
\bibitem [{\citenamefont {Srolovitz}\ \emph {et~al.}(1981)\citenamefont
  {Srolovitz}, \citenamefont {Maeda}, \citenamefont {Vitek},\ and\
  \citenamefont {Egami}}]{srolovitz_structural_1981}%
  \BibitemOpen
  \bibfield  {author} {\bibinfo {author} {\bibfnamefont {D.}~\bibnamefont
  {Srolovitz}}, \bibinfo {author} {\bibfnamefont {K.}~\bibnamefont {Maeda}},
  \bibinfo {author} {\bibfnamefont {V.}~\bibnamefont {Vitek}},\ and\ \bibinfo
  {author} {\bibfnamefont {T.}~\bibnamefont {Egami}},\ }\bibfield  {title}
  {\bibinfo {title} {Structural defects in amorphous solids {Statistical}
  analysis of a computer model},\ }\href
  {https://doi.org/10.1080/01418618108239553} {\bibfield  {journal} {\bibinfo
  {journal} {Philosophical Magazine A}\ }\textbf {\bibinfo {volume} {44}},\
  \bibinfo {pages} {847} (\bibinfo {year} {1981})}\BibitemShut {NoStop}%
\bibitem [{\citenamefont {Levashov}\ \emph {et~al.}(2008)\citenamefont
  {Levashov}, \citenamefont {Egami}, \citenamefont {Aga},\ and\ \citenamefont
  {Morris}}]{levashov_equipartition_2008}%
  \BibitemOpen
  \bibfield  {author} {\bibinfo {author} {\bibfnamefont {V.~A.}\ \bibnamefont
  {Levashov}}, \bibinfo {author} {\bibfnamefont {T.}~\bibnamefont {Egami}},
  \bibinfo {author} {\bibfnamefont {R.~S.}\ \bibnamefont {Aga}},\ and\ \bibinfo
  {author} {\bibfnamefont {J.~R.}\ \bibnamefont {Morris}},\ }\bibfield  {title}
  {\bibinfo {title} {Equipartition theorem and the dynamics of liquids},\
  }\href {https://doi.org/10.1103/PhysRevB.78.064205} {\bibfield  {journal}
  {\bibinfo  {journal} {Physical Review B}\ }\textbf {\bibinfo {volume} {78}},\
  \bibinfo {pages} {064205} (\bibinfo {year} {2008})}\BibitemShut {NoStop}%
\bibitem [{\citenamefont {Plimpton}(1995)}]{plimpton_fast_1995}%
  \BibitemOpen
  \bibfield  {author} {\bibinfo {author} {\bibfnamefont {S.}~\bibnamefont
  {Plimpton}},\ }\bibfield  {title} {\bibinfo {title} {Fast parallel algorithms
  for short-range molecular dynamics},\ }\href
  {http://www.sciencedirect.com/science/article/pii/S002199918571039X}
  {\bibfield  {journal} {\bibinfo  {journal} {Journal of computational
  physics}\ }\textbf {\bibinfo {volume} {117}},\ \bibinfo {pages} {1} (\bibinfo
  {year} {1995})}\BibitemShut {NoStop}%
\bibitem [{\citenamefont {Gale}(1997)}]{gale_gulp:_1997}%
  \BibitemOpen
  \bibfield  {author} {\bibinfo {author} {\bibfnamefont {J.~D.}\ \bibnamefont
  {Gale}},\ }\bibfield  {title} {\bibinfo {title} {{GULP}: {A} computer program
  for the symmetry-adapted simulation of solids},\ }\href
  {http://pubs.rsc.org/en/content/articlehtml/1997/ft/a606455h} {\bibfield
  {journal} {\bibinfo  {journal} {Journal of the Chemical Society, Faraday
  Transactions}\ }\textbf {\bibinfo {volume} {93}},\ \bibinfo {pages} {629}
  (\bibinfo {year} {1997})}\BibitemShut {NoStop}%
\bibitem [{\citenamefont {Keyes}(1997)}]{keyes_instantaneous_1997}%
  \BibitemOpen
  \bibfield  {author} {\bibinfo {author} {\bibfnamefont {T.}~\bibnamefont
  {Keyes}},\ }\bibfield  {title} {\bibinfo {title} {Instantaneous {Normal}
  {Mode} {Approach} to {Liquid} {State} {Dynamics}},\ }\href
  {https://doi.org/10.1021/jp963706h} {\bibfield  {journal} {\bibinfo
  {journal} {The Journal of Physical Chemistry A}\ }\textbf {\bibinfo {volume}
  {101}},\ \bibinfo {pages} {2921} (\bibinfo {year} {1997})}\BibitemShut
  {NoStop}%
\bibitem [{\citenamefont {Debenedetti}\ and\ \citenamefont
  {Stillinger}(2001)}]{debenedetti_supercooled_2001}%
  \BibitemOpen
  \bibfield  {author} {\bibinfo {author} {\bibfnamefont {P.~G.}\ \bibnamefont
  {Debenedetti}}\ and\ \bibinfo {author} {\bibfnamefont {F.~H.}\ \bibnamefont
  {Stillinger}},\ }\bibfield  {title} {\bibinfo {title} {Supercooled liquids
  and the glass transition},\ }\href
  {http://www.nature.com/nature/journal/v410/n6825/abs/410259a0.html}
  {\bibfield  {journal} {\bibinfo  {journal} {Nature}\ }\textbf {\bibinfo
  {volume} {410}},\ \bibinfo {pages} {259} (\bibinfo {year}
  {2001})}\BibitemShut {NoStop}%
\bibitem [{\citenamefont {Goldstein}(1969)}]{goldstein_viscous_1969}%
  \BibitemOpen
  \bibfield  {author} {\bibinfo {author} {\bibfnamefont {M.}~\bibnamefont
  {Goldstein}},\ }\bibfield  {title} {\bibinfo {title} {Viscous {Liquids} and
  the {Glass} {Transition}: {A} {Potential} {Energy} {Barrier} {Picture}},\
  }\href {https://doi.org/10.1063/1.1672587} {\bibfield  {journal} {\bibinfo
  {journal} {The Journal of Chemical Physics}\ }\textbf {\bibinfo {volume}
  {51}},\ \bibinfo {pages} {3728} (\bibinfo {year} {1969})}\BibitemShut
  {NoStop}%
\bibitem [{\citenamefont {Wigner}(1958)}]{wigner_distribution_1958}%
  \BibitemOpen
  \bibfield  {author} {\bibinfo {author} {\bibfnamefont {E.~P.}\ \bibnamefont
  {Wigner}},\ }\bibfield  {title} {\bibinfo {title} {On the {Distribution} of
  the {Roots} of {Certain} {Symmetric} {Matrices}},\ }\href
  {https://doi.org/10.2307/1970008} {\bibfield  {journal} {\bibinfo  {journal}
  {Annals of Mathematics}\ }\textbf {\bibinfo {volume} {67}},\ \bibinfo {pages}
  {325} (\bibinfo {year} {1958})}\BibitemShut {NoStop}%
\bibitem [{\citenamefont {Dove}(1993)}]{dove_introduction_1993}%
  \BibitemOpen
  \bibfield  {author} {\bibinfo {author} {\bibfnamefont {M.~T.}\ \bibnamefont
  {Dove}},\ }\href@noop {} {\emph {\bibinfo {title} {Introduction to lattice
  dynamics}}},\ Vol.~\bibinfo {volume} {4}\ (\bibinfo  {publisher} {Cambridge
  university press},\ \bibinfo {year} {1993})\BibitemShut {NoStop}%
\bibitem [{\citenamefont {Rahman}(1964)}]{rahman_correlations_1964}%
  \BibitemOpen
  \bibfield  {author} {\bibinfo {author} {\bibfnamefont {A.}~\bibnamefont
  {Rahman}},\ }\bibfield  {title} {\bibinfo {title} {Correlations in the
  {Motion} of {Atoms} in {Liquid} {Argon}},\ }\href
  {https://doi.org/10.1103/PhysRev.136.A405} {\bibfield  {journal} {\bibinfo
  {journal} {Physical Review}\ }\textbf {\bibinfo {volume} {136}},\ \bibinfo
  {pages} {A405} (\bibinfo {year} {1964})}\BibitemShut {NoStop}%
\bibitem [{\citenamefont {Verkerk}\ \emph {et~al.}(1989)\citenamefont
  {Verkerk}, \citenamefont {Westerweel}, \citenamefont {Bafile}, \citenamefont
  {de~Graaf}, \citenamefont {Montfrooij},\ and\ \citenamefont
  {de~Schepper}}]{verkerk_velocity_1989}%
  \BibitemOpen
  \bibfield  {author} {\bibinfo {author} {\bibfnamefont {P.}~\bibnamefont
  {Verkerk}}, \bibinfo {author} {\bibfnamefont {J.}~\bibnamefont {Westerweel}},
  \bibinfo {author} {\bibfnamefont {U.}~\bibnamefont {Bafile}}, \bibinfo
  {author} {\bibfnamefont {L.~A.}\ \bibnamefont {de~Graaf}}, \bibinfo {author}
  {\bibfnamefont {W.}~\bibnamefont {Montfrooij}},\ and\ \bibinfo {author}
  {\bibfnamefont {I.~M.}\ \bibnamefont {de~Schepper}},\ }\bibfield  {title}
  {\bibinfo {title} {Velocity autocorrelation function of dense hydrogen gas
  determined by neutron scattering},\ }\href
  {https://doi.org/10.1103/PhysRevA.40.2860} {\bibfield  {journal} {\bibinfo
  {journal} {Physical Review A}\ }\textbf {\bibinfo {volume} {40}},\ \bibinfo
  {pages} {2860} (\bibinfo {year} {1989})}\BibitemShut {NoStop}%
\bibitem [{\citenamefont {Stassen}\ and\ \citenamefont
  {Gburski}(1994)}]{stassen_instantaneous_1994}%
  \BibitemOpen
  \bibfield  {author} {\bibinfo {author} {\bibfnamefont {H.}~\bibnamefont
  {Stassen}}\ and\ \bibinfo {author} {\bibfnamefont {Z.~E.}\ \bibnamefont
  {Gburski}},\ }\bibfield  {title} {\bibinfo {title} {Instantaneous normal mode
  analysis of binary liquid {Ar}-{Kr} mixtures},\ }\href
  {https://doi.org/10.1016/0009-2614(93)E1390-3} {\bibfield  {journal}
  {\bibinfo  {journal} {Chemical Physics Letters}\ }\textbf {\bibinfo {volume}
  {217}},\ \bibinfo {pages} {325} (\bibinfo {year} {1994})}\BibitemShut
  {NoStop}%
\bibitem [{\citenamefont {Melzer}\ \emph {et~al.}(2012)\citenamefont {Melzer},
  \citenamefont {Schella}, \citenamefont {Schablinski}, \citenamefont {Block},\
  and\ \citenamefont {Piel}}]{melzer_instantaneous_2012}%
  \BibitemOpen
  \bibfield  {author} {\bibinfo {author} {\bibfnamefont {A.}~\bibnamefont
  {Melzer}}, \bibinfo {author} {\bibfnamefont {A.}~\bibnamefont {Schella}},
  \bibinfo {author} {\bibfnamefont {J.}~\bibnamefont {Schablinski}}, \bibinfo
  {author} {\bibfnamefont {D.}~\bibnamefont {Block}},\ and\ \bibinfo {author}
  {\bibfnamefont {A.}~\bibnamefont {Piel}},\ }\bibfield  {title} {\bibinfo
  {title} {Instantaneous {Normal} {Mode} {Analysis} of {Melting} of {Finite}
  {Dust} {Clusters}},\ }\href@noop {} {\bibfield  {journal} {\bibinfo
  {journal} {Physical Review Letters}\ ,\ \bibinfo {pages} {5}} (\bibinfo
  {year} {2012})}\BibitemShut {NoStop}%
\bibitem [{\citenamefont {Liu}\ \emph {et~al.}(2012)\citenamefont {Liu},
  \citenamefont {Shi}, \citenamefont {Xu}, \citenamefont {Zhang}, \citenamefont
  {Zhang}, \citenamefont {Chen}, \citenamefont {Li}, \citenamefont {Uher},
  \citenamefont {Day},\ and\ \citenamefont {Snyder}}]{liu_copper_2012}%
  \BibitemOpen
  \bibfield  {author} {\bibinfo {author} {\bibfnamefont {H.}~\bibnamefont
  {Liu}}, \bibinfo {author} {\bibfnamefont {X.}~\bibnamefont {Shi}}, \bibinfo
  {author} {\bibfnamefont {F.}~\bibnamefont {Xu}}, \bibinfo {author}
  {\bibfnamefont {L.}~\bibnamefont {Zhang}}, \bibinfo {author} {\bibfnamefont
  {W.}~\bibnamefont {Zhang}}, \bibinfo {author} {\bibfnamefont
  {L.}~\bibnamefont {Chen}}, \bibinfo {author} {\bibfnamefont {Q.}~\bibnamefont
  {Li}}, \bibinfo {author} {\bibfnamefont {C.}~\bibnamefont {Uher}}, \bibinfo
  {author} {\bibfnamefont {T.}~\bibnamefont {Day}},\ and\ \bibinfo {author}
  {\bibfnamefont {G.~J.}\ \bibnamefont {Snyder}},\ }\bibfield  {title}
  {\bibinfo {title} {Copper ion liquid-like thermoelectrics},\ }\href
  {https://doi.org/10.1038/nmat3273} {\bibfield  {journal} {\bibinfo  {journal}
  {Nature Materials}\ }\textbf {\bibinfo {volume} {11}},\ \bibinfo {pages}
  {422} (\bibinfo {year} {2012})}\BibitemShut {NoStop}%
\bibitem [{\citenamefont {Voneshen}\ \emph {et~al.}(2017)\citenamefont
  {Voneshen}, \citenamefont {Walker}, \citenamefont {Refson},\ and\
  \citenamefont {Goff}}]{voneshen_hopping_2017}%
  \BibitemOpen
  \bibfield  {author} {\bibinfo {author} {\bibfnamefont {D.}~\bibnamefont
  {Voneshen}}, \bibinfo {author} {\bibfnamefont {H.}~\bibnamefont {Walker}},
  \bibinfo {author} {\bibfnamefont {K.}~\bibnamefont {Refson}},\ and\ \bibinfo
  {author} {\bibfnamefont {J.}~\bibnamefont {Goff}},\ }\bibfield  {title}
  {\bibinfo {title} {Hopping {Time} {Scales} and the {Phonon}-{Liquid}
  {Electron}-{Crystal} {Picture} in {Thermoelectric} {Copper} {Selenide}},\
  }\href {https://doi.org/10.1103/PhysRevLett.118.145901} {\bibfield  {journal}
  {\bibinfo  {journal} {Physical Review Letters}\ }\textbf {\bibinfo {volume}
  {118}},\ \bibinfo {pages} {145901} (\bibinfo {year} {2017})}\BibitemShut
  {NoStop}%
\bibitem [{\citenamefont {Niedziela}\ \emph {et~al.}(2019)\citenamefont
  {Niedziela}, \citenamefont {Bansal}, \citenamefont {May}, \citenamefont
  {Ding}, \citenamefont {Lanigan-Atkins}, \citenamefont {Ehlers}, \citenamefont
  {Abernathy}, \citenamefont {Said},\ and\ \citenamefont
  {Delaire}}]{niedziela_selective_2019}%
  \BibitemOpen
  \bibfield  {author} {\bibinfo {author} {\bibfnamefont {J.~L.}\ \bibnamefont
  {Niedziela}}, \bibinfo {author} {\bibfnamefont {D.}~\bibnamefont {Bansal}},
  \bibinfo {author} {\bibfnamefont {A.~F.}\ \bibnamefont {May}}, \bibinfo
  {author} {\bibfnamefont {J.}~\bibnamefont {Ding}}, \bibinfo {author}
  {\bibfnamefont {T.}~\bibnamefont {Lanigan-Atkins}}, \bibinfo {author}
  {\bibfnamefont {G.}~\bibnamefont {Ehlers}}, \bibinfo {author} {\bibfnamefont
  {D.~L.}\ \bibnamefont {Abernathy}}, \bibinfo {author} {\bibfnamefont
  {A.}~\bibnamefont {Said}},\ and\ \bibinfo {author} {\bibfnamefont
  {O.}~\bibnamefont {Delaire}},\ }\bibfield  {title} {\bibinfo {title}
  {Selective breakdown of phonon quasiparticles across superionic transition in
  {CuCrSe2}},\ }\href {https://doi.org/10.1038/s41567-018-0298-2} {\bibfield
  {journal} {\bibinfo  {journal} {Nature Physics}\ }\textbf {\bibinfo {volume}
  {15}},\ \bibinfo {pages} {73} (\bibinfo {year} {2019})}\BibitemShut {NoStop}%
\bibitem [{\citenamefont {Seeley}\ \emph {et~al.}(1991)\citenamefont {Seeley},
  \citenamefont {Keyes},\ and\ \citenamefont {Madan}}]{seeley_isobaric_1991}%
  \BibitemOpen
  \bibfield  {author} {\bibinfo {author} {\bibfnamefont {G.}~\bibnamefont
  {Seeley}}, \bibinfo {author} {\bibfnamefont {T.}~\bibnamefont {Keyes}},\ and\
  \bibinfo {author} {\bibfnamefont {B.}~\bibnamefont {Madan}},\ }\bibfield
  {title} {\bibinfo {title} {Isobaric diffusion constants in simple liquids and
  normal mode analysis},\ }\href {https://doi.org/10.1063/1.460787} {\bibfield
  {journal} {\bibinfo  {journal} {The Journal of Chemical Physics}\ }\textbf
  {\bibinfo {volume} {95}},\ \bibinfo {pages} {3847} (\bibinfo {year}
  {1991})}\BibitemShut {NoStop}%
\bibitem [{\citenamefont {Nave}\ \emph {et~al.}(2000)\citenamefont {Nave},
  \citenamefont {Scala}, \citenamefont {Starr}, \citenamefont {Sciortino},\
  and\ \citenamefont {Stanley}}]{nave_instantaneous_2000}%
  \BibitemOpen
  \bibfield  {author} {\bibinfo {author} {\bibfnamefont {E.~L.}\ \bibnamefont
  {Nave}}, \bibinfo {author} {\bibfnamefont {A.}~\bibnamefont {Scala}},
  \bibinfo {author} {\bibfnamefont {F.~W.}\ \bibnamefont {Starr}}, \bibinfo
  {author} {\bibfnamefont {F.}~\bibnamefont {Sciortino}},\ and\ \bibinfo
  {author} {\bibfnamefont {H.~E.}\ \bibnamefont {Stanley}},\ }\bibfield
  {title} {\bibinfo {title} {Instantaneous {Normal} {Mode} {Analysis} of
  {Supercooled} {Water}},\ }\href@noop {} {\bibfield  {journal} {\bibinfo
  {journal} {Physical Review Letters}\ }\textbf {\bibinfo {volume} {84}},\
  \bibinfo {pages} {4} (\bibinfo {year} {2000})}\BibitemShut {NoStop}%
\bibitem [{\citenamefont {Clapa}\ \emph {et~al.}(2012)\citenamefont {Clapa},
  \citenamefont {Kottos},\ and\ \citenamefont
  {Starr}}]{clapa_localization_2012}%
  \BibitemOpen
  \bibfield  {author} {\bibinfo {author} {\bibfnamefont {V.~I.}\ \bibnamefont
  {Clapa}}, \bibinfo {author} {\bibfnamefont {T.}~\bibnamefont {Kottos}},\ and\
  \bibinfo {author} {\bibfnamefont {F.~W.}\ \bibnamefont {Starr}},\ }\bibfield
  {title} {\bibinfo {title} {Localization transition of instantaneous normal
  modes and liquid diffusion},\ }\href {https://doi.org/10.1063/1.3701564}
  {\bibfield  {journal} {\bibinfo  {journal} {The Journal of Chemical Physics}\
  }\textbf {\bibinfo {volume} {136}},\ \bibinfo {pages} {144504} (\bibinfo
  {year} {2012})}\BibitemShut {NoStop}%
\bibitem [{\citenamefont {Schirmacher}\ \emph {et~al.}(2022)\citenamefont
  {Schirmacher}, \citenamefont {Bryk},\ and\ \citenamefont
  {Ruocco}}]{schirmacher_modeling_2022}%
  \BibitemOpen
  \bibfield  {author} {\bibinfo {author} {\bibfnamefont {W.}~\bibnamefont
  {Schirmacher}}, \bibinfo {author} {\bibfnamefont {T.}~\bibnamefont {Bryk}},\
  and\ \bibinfo {author} {\bibfnamefont {G.}~\bibnamefont {Ruocco}},\
  }\bibfield  {title} {\bibinfo {title} {Modeling the instantaneous normal mode
  spectra of liquids as that of unstable elastic media},\ }\href
  {https://doi.org/10.1073/pnas.2119288119} {\bibfield  {journal} {\bibinfo
  {journal} {Proceedings of the National Academy of Sciences}\ }\textbf
  {\bibinfo {volume} {119}},\ \bibinfo {pages} {e2119288119} (\bibinfo {year}
  {2022})}\BibitemShut {NoStop}%
\end{thebibliography}%


%
\end{document}


\title{Supplementary Materials for\\
Microscopic view of heat capacity of matter: solid, liquid, and gas}

\author{Jaeyun Moon}
\email{To whom correspondence should be addressed; E-mail:  moonj@ornl.gov}
 \affiliation{Materials Science and Technology Division,\\ Oak Ridge National Laboratory, Oak Ridge, Tennessee 37831, USA}
\author{Simon Th\'ebaud}
 \affiliation{Materials Science and Technology Division,\\ Oak Ridge National Laboratory, Oak Ridge, Tennessee 37831, USA}
\author{Lucas Lindsay}
 \affiliation{Materials Science and Technology Division,\\ Oak Ridge National Laboratory, Oak Ridge, Tennessee 37831, USA}
 
\author{Takeshi Egami}
\affiliation{Materials Science and Technology Division,\\ Oak Ridge National Laboratory, Oak Ridge, Tennessee 37831, USA\\
 Department of Materials Science and Engineering,\\ University of Tennessee, Knoxville, Tennessee 37996, USA\\
 Department of Physics and Astronomy,\\ University of Tennessee, Knoxville, Tennessee 37996, USA}

\date{\today}

\maketitle
\clearpage

\renewcommand{\figurename}{Figure}
\renewcommand{\thefigure}{S\arabic{figure}}
\setcounter{figure}{0}

\section{Simulation details}

For argon, FCC structure of 1372 atoms with a lattice parameter of 5.2686 \AA \, ($\rho_0$) was used as an initial input structure for molecular dynamics simulations. Temperatures considered were from 1 to $10^8$ K in increments of factors of 10 (log scale) excluding 100 K near the melting temperature. Extremely high temperatures were necessary to reach the gas limit ($C_V = 1.5Nk_B$). MD Timesteps were 0.1 and 0.01 fs depending on the temperature to capture the fast atomic dynamics. For silicon, FCC structure with 1728 atoms with lattice parameter of 5.431 \AA \, ($\rho_0$) was employed. Calculations were done at temperatures from 1 to $10^7$ K with varying timesteps of 0.5 and 0.05 fs. For iron, BCC structure with 2000 atoms with a lattice parameter of 2.867 \AA \, ($\rho_0$) was studied from 1 to $10^6$ K. Timesteps of 0.1 and 0.01 fs were used.  At each temperature, all systems were equilibrated for $5 \times 10^6$ timesteps in the canonical ensemble (NVT) prior to data recording of another $5 \times 10^6$ timesteps under the same ensemble. Relativistic corrections were not needed at high temperatures due to relatively heavy masses of our systems. 

These systems have different atomic structures in the solid state: face centered cubic (argon), face centered cubic (silicon), and body centered cubic structures (iron) below melting temperatures under atmospheric pressures. Upon melting, liquid argon and iron atoms are close packed with coordination numbers around 13 to 14. In comparison, silicon possesses more complex temperature dependent structural features. Due to formation of metallic bonds upon melting, the coordination number of silicon increases from 4 to 6 or more and there also exists a low density liquid to high density liquid transition, commonly observed in other tetrahedral systems including water. Both structural changes in silicon have been well-characterized by the Stillinger-Weber potential used here. Having the various structural features and bond natures described above, argon, silicon, and iron are versatile test systems for the purpose of this study.

\section{Diffusion coefficients and instability parameters}

\begin{figure}
	\centering
	\includegraphics[width=0.7\linewidth]{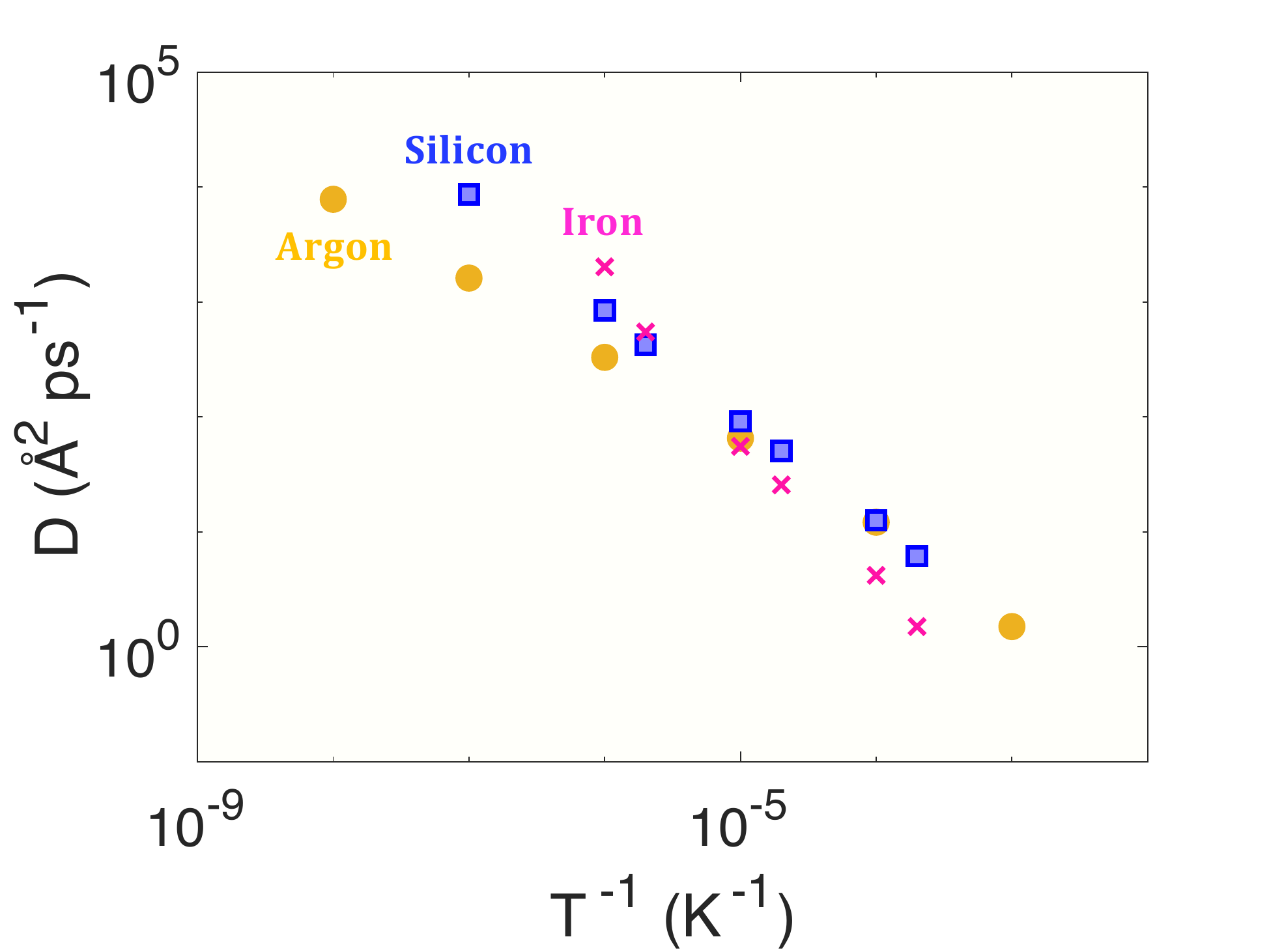}
	\caption{Temperature dependent self-diffusion coefficients, $D$, for all $\rho_0$ systems. Yellow circles, blue squares, and pink crosses are for argon, silicon, and iron, respectively. Monotonic increase in $D$ with increase in temperature is observed.}
	\label{fig:Diffusion}
\end{figure}

Temperature dependent self-diffusion coefficients, $D$, for all $\rho_0$ systems were calculated by 
\begin{equation}
    D = \lim_{t \to \infty}\frac{1}{6t}\langle [\boldsymbol{r}_i(t) -  \boldsymbol{r}_i(0)]^2 \rangle 
\end{equation}
where $t$ is time and $\boldsymbol{r}_i(t)$ is the time dependent position of atom $i$. As shown in Fig. \ref{fig:Diffusion}, we observe a continuous increase in $D$ for all systems with increase in temperature. Error bars are smaller than the symbol sizes. 

Prior works have attempted to relate self-diffusion coefficients, $D$, of liquids to the fraction of imaginary modes over the total number of modes, $3N$ \cite{seeley_isobaric_1991,nave_instantaneous_2000,clapa_localization_2012}. Some have argued that $D$ is directly proportional to this fraction \cite{seeley_isobaric_1991} while others have argued for more subtle relations between $D$ and the fraction of `delocalized' imaginary modes \cite{clapa_localization_2012}. However, INM($\omega$) for our high temperature liquid and gas systems challenge these ideas: (1) we see a dramatic slow-down of how fast the density of states changes with temperature as evident in Fig. 2 (A) in the main manuscript where temperature increases evenly by a factor of 10 but self-diffusion coefficients at these temperatures continue to increase with temperature as shown in Fig. S1. (2) All imaginary modes in gas states at high temperatures are localized as confirmed by their inverse participation ratios (not shown here). More recent work has emphasized the importance of relative shapes of real and negative eigenvalue distributions of the dynamical matrices in understanding the nature of vibrational modes in liquids and how they become more symmetric with increase in temperature in liquids above glass transition temperatures \cite{schirmacher_modeling_2022}. 

\begin{figure}[h!]
	\centering
	\includegraphics[width=0.7\linewidth]{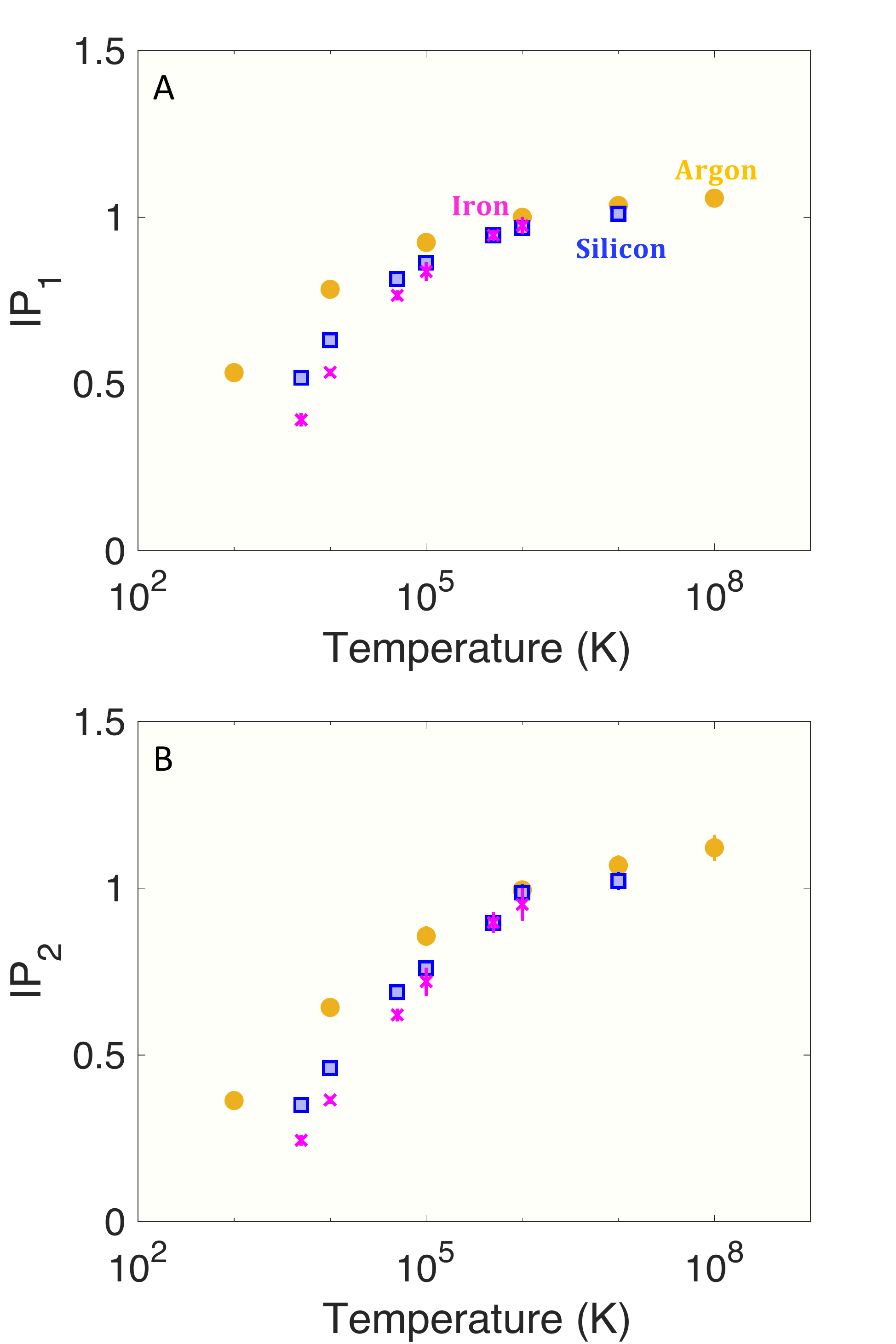}
	\caption{Temperature dependent instability parameter for all $\rho_0$ systems. Yellow circles, blue squares, and pink crosses are for argon, silicon, and iron, respectively. }
	\label{fig:IP}
\end{figure}

We have introduced two parameters describing gasness of a system known as instability parameters given by $\text{IP}_1 = \frac{2N_i}{3N}$ and $\text{IP}_2 = \frac{N_i}{3N-N_i}$, where $N_i$ is the total number of imaginary modes. Temperature dependence of these parameters are shown in Fig. \ref{fig:IP}. We observe a clear slowdown in increase in both instability parameters with temperature around $\text{IP}_{1,2} = 1$. Instability parameters are slightly above 1 at high temperatures. We believe that this is due to small statistics arising from using relatively small system sizes. To adequately test this issue, we would need to do lattice dynamics calculations on much bigger systems of $N = 10^4$ to $N = 10^6$ atoms; however, diagonalizing such large $3N$ by $3N$ matrices is challenging. We have additionally done lattice dynamics calculations of argon gas at a much lower density of $0.1 \rho_0$ from 500 K to $10^4$ K and $\text{IP}_{1,2}$ were constant at $\sim 1.1$.

\clearpage

\section{INM($\omega$) and VACF($\omega$) equivalence for solids}

\begin{figure}[h!]
	\centering
	\includegraphics[width=1\linewidth]{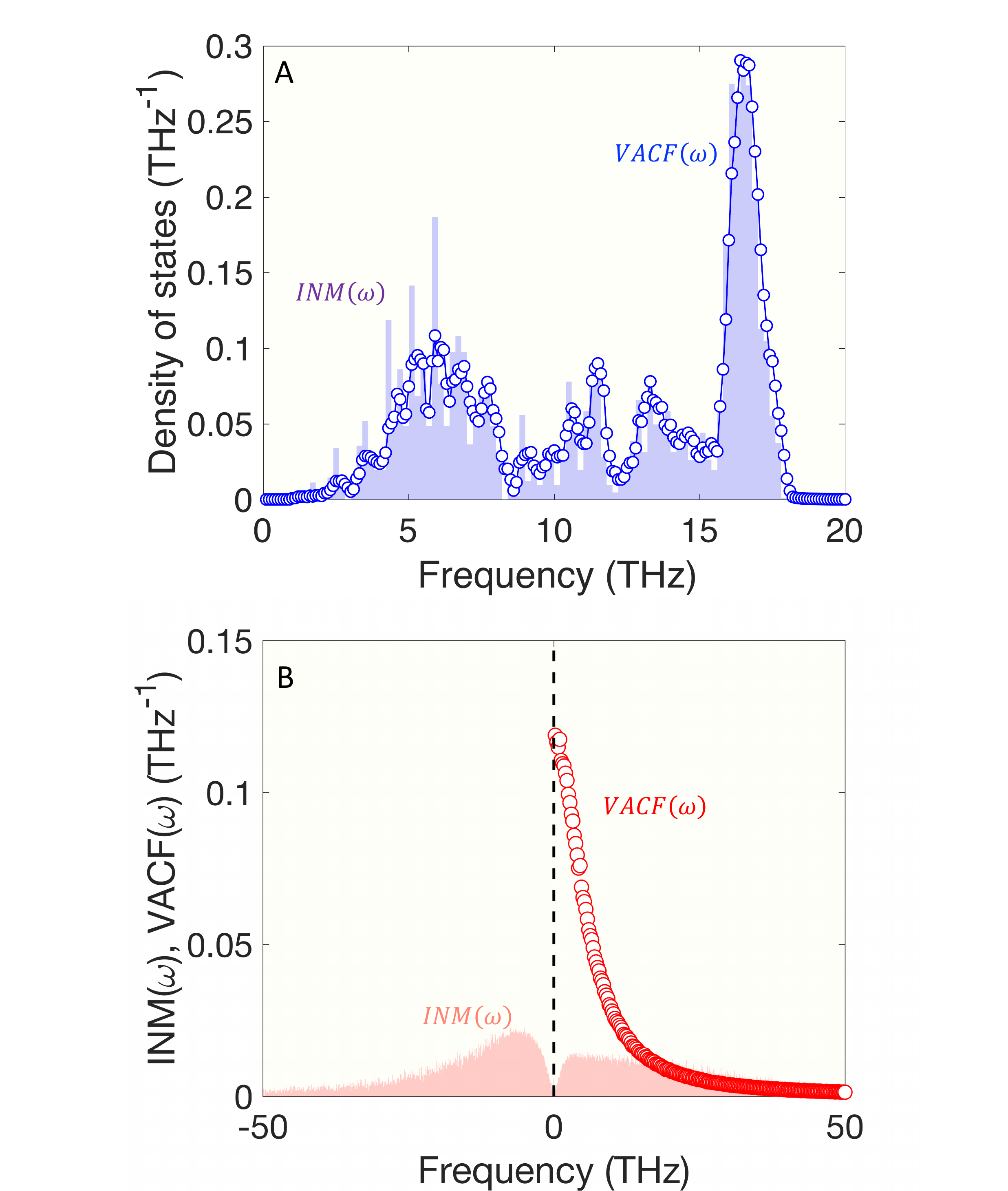}
	\caption{(A) Phonon density of states for crystalline silicon at 1 K. We see nearly identical spectra between INM($\omega$) and VACF($\omega$), as expected. (B) INM($\omega$) and VACF($\omega$) of silicon gas at $10^7$ K. The Stillinger-Weber potential was used. Negative frequencies denote modes with imaginary frequencies.}
	\label{fig:DOS}
\end{figure}

Instantaneous normal mode and velocity autocorrelation spectra for solids are known to be equivalent under harmonic approximations and are both used to obtain phonon density of states of solids. An example of this equivalence is shown in Fig. \ref{fig:DOS}(A) for crystalline silicon at 1 K with the Stillinger-Weber potential by treating the entire domain as a unit cell and only the $\Gamma$ point is considered. On the other hand, strikingly different spectra are observed for liquid and gas systems as demonstrated in Fig. \ref{fig:DOS}(B). In this work, we make connections between these spectra via instability parameters characterizing the nature of the atomic degrees of freedom. 
\clearpage 

\clearpage
\section{Temperature dependent specific heat}

\begin{figure}
	\centering
	\includegraphics[width=0.5\linewidth]{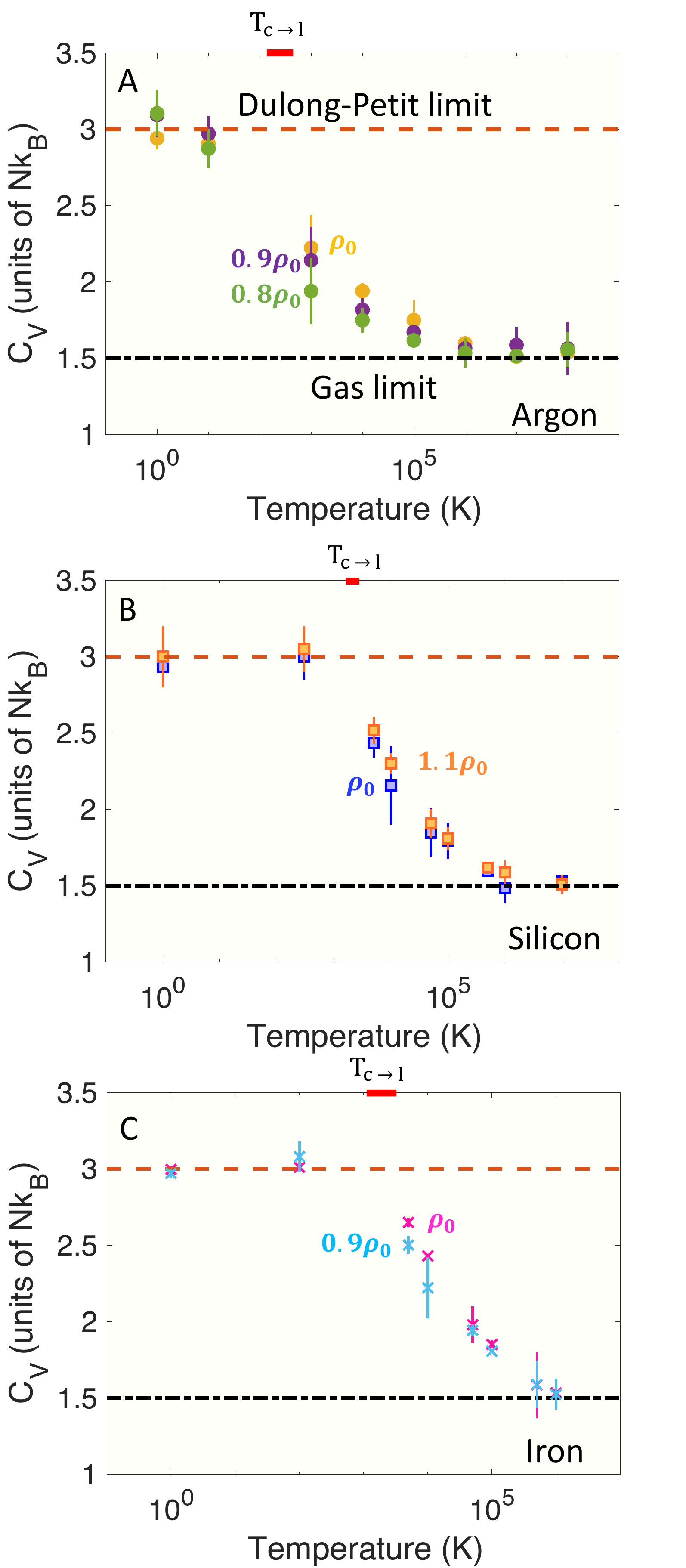}
	\caption{Temperature dependent constant volume heat capacity of (A) argon, (B) silicon, and (C) iron with different densities. Different marker colors represent different densities. Brown dash lines and black dash-dot lines represent Dulong-Petit limits and gas-limits for monatomic systems. $T_{c \: \rightarrow \: l}$ marked by red lines mean crystal to liquid melting temperatures.}
	\label{fig:Cv_T}
\end{figure}

For canonical ensembles in equilibrium, statistical mechanics dictates that specific heat is related to energy fluctuations by 
\begin{equation}
    C_V = \frac{\langle E \rangle^2 - \langle E^2 \rangle}{k_BT^2}
\end{equation}
where angled brackets represent ensemble averages. Temperature dependent specific heats of all systems studied in this work are shown in Fig. \ref{fig:Cv_T}. Melting temperatures are denoted by the red lines on the $x$-axis. For molecular dynamics simulations, which are classical, specific heats for solids even at low temperatures are $3Nk_B$ as observed in the figure. We observe clear transitions to the gas limit as we increase the temperature. 

\begin{figure}[h!]
	\centering
	\includegraphics[width=0.7\linewidth]{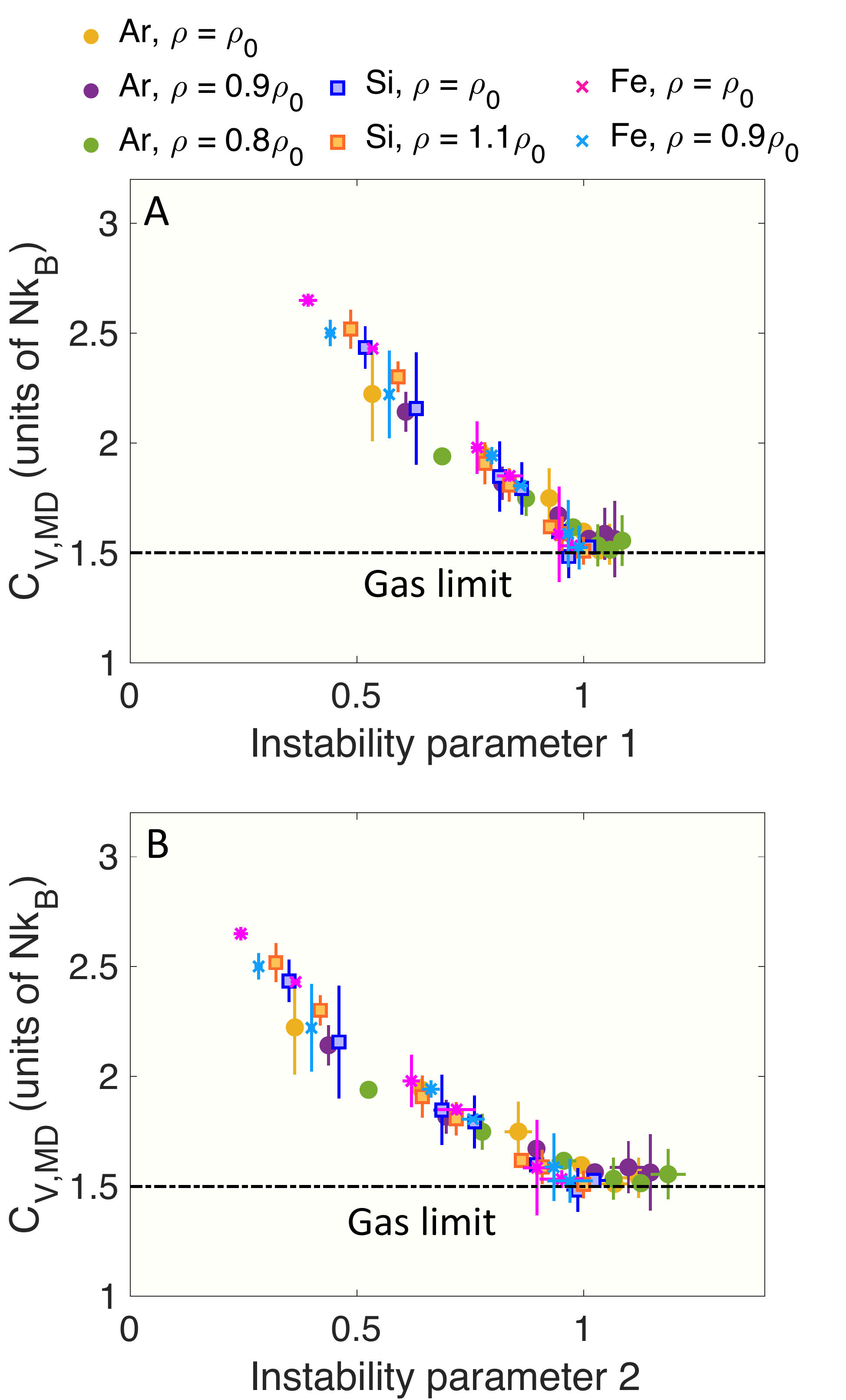}
	\caption{Instability parameters versus constant volume specific heats.}
	\label{fig:IP_Cv}
\end{figure}

In this work, we propose to understand heat capacity of liquids via instability parameters. Instability parameters versus constant volume specific heats are shown in Fig. \ref{fig:IP_Cv}. Despite different spectral shapes and features among argon, silicon, and iron systems as demonstrated in Fig. 2 and Fig. 3, strong correlations between instability parameters and heat capacities with small spread are evident. As predicted, specific heat goes to the gas limit of $1.5Nk_B$ near IP\textsubscript{1,2} = 1.

\clearpage

\clearpage
